\documentclass[11pt,a4paper]{article}

\usepackage{jheppub} 

\newcommand{\gR}{g_\mathrm{R}}
\newcommand{\mre}{\mathrm{e}}
\newcommand{\mrO}{\mathrm{O}}
\newcommand{\mrd}{\mathrm{d}}
\newcommand{\cO}{\mathcal{O}}
\newcommand{\evec}{\vec{e}} 
\newcommand{\Jvec}{\vec{J}} 

\newcommand{\Z}{{\mathbb{Z}}}

\newcommand{\CP}{{\mathbb{C}\mathrm{P}}}

\newcommand{\p}{\partial}

\title{Drastic Reduction of Cutoff Effects \\
in 2-d Lattice \boldmath $\mrO(N)$ Models}

\author[a]{J.\ Balog,}
\author[b,c]{F.\ Niedermayer,}
\author[d,b]{M.\ Pepe,}
\author[e]{P.\ Weisz,}
\author[b]{and U.-J.\ Wiese}

\affiliation[a]{Institute for Particle and Nuclear Physics, 
Wigner Research Centre for Physics, \\
MTA Lend\"ulet Holographic QFT Group,
1525 Budapest 114, P.O.B.\ 49, Hungary}
\affiliation[b]{Albert Einstein Center for Fundamental Physics \\
Institute for Theoretical Physics,
Bern University, Sidlerstr.\ 5, 3012 Bern, Switzerland}
\affiliation[c]{Institute for Theoretical Physics -- HAS,
E\"otv\"os University, \\
P\'azm\'any s\'et\'any 1/a, 1117 Budapest, Hungary}
\affiliation[d]{INFN, Istituto Nazionale di Fisica Nucleare \\ 
Sezione di Milano-Bicocca,
Edificio U2, Piazza della Scienza 3 - 20126 Milano, Italy}
\affiliation[e]{Max-Planck-Institut f\"ur Physik, 80805 Munich, Germany}

\emailAdd{balog.janos@wigner.mta.hu}
\emailAdd{niedermayer@itp.unibe.ch}
\emailAdd{Michele.Pepe@mib.infn.it}
\emailAdd{pew@mpp.mpg.de}
\emailAdd{wiese@itp.unibe.ch}

\abstract{We investigate the cutoff effects in 2-d lattice $\mrO(N)$ models 
  for a variety of lattice actions, and we identify a class of very simple 
  actions for which the lattice artifacts are extremely small. One action 
  agrees with the standard action, except that it constrains neighboring 
  spins to a maximal relative angle $\delta$. We fix $\delta$ by demanding 
  that a particular value of the step scaling function agrees with its 
  continuum result already on a rather coarse lattice. Remarkably, the 
  cutoff effects of the entire step scaling function are then reduced to 
  the per mille level. This also applies to the $\theta$-vacuum effects 
  of the step scaling function in the 2-d $\mrO(3)$ model. The cutoff effects 
  of other physical observables including the renormalized coupling $\gR$ 
  and the mass in the isotensor channel are also reduced drastically. 
  Another choice, the mixed action, which combines the standard quadratic 
  with an appropriately tuned large quartic term, also has extremely small 
  cutoff effects. The size of cutoff effects is also investigated 
  analytically in 1-d and at $N = \infty$ in 2-d.}

\subheader{\hfill MPP-2012-125}

\begin{document} 

\maketitle

\section{Introduction}
\label{Introduction}

2-d $\mrO(N)$ models share many features with 4-d non-Abelian gauge theories.
They are asymptotically free, have a non-perturbatively generated mass gap, 
and, for $N = 3$, even instantons and $\theta$-vacuum effects. 
2-d $\mrO(N)$ models are integrable and have an analytically known 
exact S-matrix \cite{Zam79,Pol83,Wie85}. 
Based on this result, the exact mass gap has been extracted 
analytically \cite{Has90}. 
This has even been extended to the mass gap $m(L)$ in a finite periodic 
volume of size $L$ \cite{Bal04,Heg05,Bal05,Bal10}, 
which then provides exact information on the step scaling function 
introduced in \cite{Lue91}. 
Furthermore, $\mrO(N)$ models can be simulated very efficiently with 
the Wolff cluster algorithm \cite{Wol89,Wol90}. For these reasons, 2-d $\mrO(N)$
models are ideally suited as toy models for QCD, on which non-perturbative 
lattice methods can be tested with exquisite precision. Systematically 
controlling ultra-violet cutoff effects due to a finite lattice spacing $a$ is 
a major objective of any lattice calculation. Symanzik's improvement program 
provides a reliable effective field theory basis for achieving this goal
\cite{Sym83,Sym83a,Lue85,Lue85a}. Interestingly, for many years the observed 
cutoff effects of the step scaling function in the 2-d $\mrO(3)$ model, which 
seemed to be of order $\cO(a)$, were in apparent contradiction with the 
$\cO(a^2)$ behavior predicted by Symanzik's theory \cite{Has02}. 
Only recently, a careful higher-order investigation of Symanzik's effective 
theory resolved this puzzle by identifying large logarithmic corrections to the 
$\cO(a^2)$ effects, which mimic $\cO(a)$ behavior \cite{Bal09}.

While Symanzik's improvement program aims at reducing cutoff effects in a 
systematic manner, order by order in the lattice spacing, the perfect action 
approach aims at completely eliminating cutoff effects at least at the 
classical level \cite{Has94}. The fixed point action corresponding to a given 
renormalization group blocking transformation is indeed a classically perfect 
action, which is completely free of lattice spacing artifacts, even at 
arbitrarily coarse lattices. Remarkably, a practical parametrization of the 
classically perfect fixed point action, which includes a large number of terms 
beyond the standard nearest-neighbor coupling, was found to drastically reduce 
cutoff effects even at the quantum level \cite{Has94}. Recently, the study of 
cutoff effects in the 2-d $\mrO(3)$ model has been driven to another 
extreme by studying topological actions \cite{Bie10}. 
Topological lattice actions are invariant against small continuous 
deformations of the lattice fields. The simplest topological action 
constrains the relative angle of neighboring $\mrO(N)$ 
spins to a maximally allowed angle $\delta$, and assigns zero action to all 
allowed configurations. This action does not have the correct classical 
continuum limit, it cannot be studied in perturbation theory, and it violates 
the Schwarz inequality between action and topological charge for $N = 3$. 
Hence, one may consider this action as tree-level impaired (rather than 
e.g.~1-loop Symanzik improved). 
Despite these deficiencies, the 2-d $\mrO(3)$ model with a topological lattice 
action was still found to have the correct quantum continuum 
limit \cite{Bie10}. 
Its cutoff effects at practically accessible correlation lengths
were found to be even smaller than those of the standard action.
Interestingly, the topological lattice action approaches the 
continuum limit of the step scaling function from below, while the standard 
action approaches it from above.

In this paper, we combine the standard and the topological action to a 
non-topological constrained action. The relative angle between neighboring spins
is again limited by a maximal angle $\delta$, but allowed configurations are
now assigned the standard action value. Similar actions with a constraint have 
been used before in various contexts 
\cite{Lue82,Pat92,Pat93,Aiz94,Bie95,Has96,
Her99,Lue99,Lue00,Pat02,Fuk03,Fuk04,Fuk06,Jan06}. 
Here we optimize the constraint angle $\delta$ to reduce the cutoff effects 
in 2-d $\mrO(N)$ models. 
Remarkably, the cutoff effects of a variety of physical quantities including 
the step scaling function, the renormalized coupling $\gR$, and the mass in the 
isotensor channel, as well as the vacuum angle $\theta$-dependence of 
the mass gap for $N = 3$, turn out to be at most a few per mille, 
even on rather coarse lattices. 
This provides us with a very simple nearest-neighbor action that can 
be simulated very efficiently with the Wolff cluster algorithm 
\cite{Wol89,Wol90}. In fact, the optimized constrained action reduces cutoff 
effects at least to the same extent as the parametrized classically perfect 
action, but is a lot simpler. A mixed action, which combines the standard
quadratic with a large quartic term suppresses cutoff effects equally well.

The paper is organized as follows. In section~2 we introduce the various actions
to be considered in this work. Section~3 contains the investigation of the
corresponding cutoff effects in the 2-d $\mrO(3)$ model first at vacuum angle 
$\theta = 0$, and then also at $\theta \neq 0$. In section~4, we study the 2-d 
$\mrO(4)$ and $\mrO(8)$ models in a similar manner, concentrating on 
the step-scaling functions. 
Section~5 addresses the cutoff effects in the $N = \infty$ limit 
analytically. Finally, Section~6 contains our conclusions. 
An analytic investigation of cutoff effects in the 1-d $\mrO(3)$ model is 
relegated to appendix~A, while some technical details of the $N = \infty$ 
calculation are presented in appendix~B.

\section{Lattice Actions for \boldmath $\mrO(N)$ Models}
\label{LatticeAct}

In this section we introduce various actions for $\mrO(N)$ models. In later 
sections we will compare the corresponding cutoff effects in order to identify 
a highly optimized lattice action. In the continuum, the action of the 2-d 
$\mrO(N)$ model is given by
\begin{equation}
  \label{Scont}
  S[\vec e] = \frac{1}{2 g^2} \int \mrd^2x \ \p_\mu \vec e \cdot \p_\mu \vec e,
\end{equation}
where $\vec e(x) = (e_1(x),e_2(x),\dots,e_N(x))$ is an $N$-component unit-vector
field, and $g$ is the dimensionless coupling constant. Just as non-Abelian gauge
theories in 4 space-time dimensions, 2-d $\mrO(N)$ models (with $N > 2$) are
asymptotically free and have a non-perturbatively generated mass gap. 
For $N = 3$ one can define the topological charge
\begin{equation}
  \label{Qcont}
  Q[\vec e] =  \frac{1}{8 \pi} \int \mrd^2x \ \varepsilon_{\mu\nu}
  \vec e \cdot (\p_\mu \vec e \times \p_\nu \vec e),
\end{equation}
which is an element of the homotopy group $\Pi_2[S^2] = \Z$. In that case, one 
can introduce a vacuum angle $\theta$ and add $i \theta Q[\vec e]$ to the
Euclidean action. Interestingly, as we have recently demonstrated, $\theta$ is
a relevant parameter and there are distinct continuum theories for each value
$\theta \in [0,\pi]$ \cite{Boe11}. This conclusion has been further 
supported by \cite{Nog12}. For $N > 3$, on the other hand, 2-d $\mrO(N)$ models 
are topologically trivial.

The standard action for the $\mrO(N)$ model on a 2-d square space-time lattice 
takes the form
\begin{equation}
  \label{Sstd}
  S_{\mathrm{std}}[\vec e] = \beta \sum_{x,\mu} 
  (1 - \vec e_x \cdot \vec e_{x+\hat\mu}),
\end{equation}
where $\vec e_x$ is an $N$-component unit-vector associated with the lattice 
site $x$, and $\hat\mu$ points from $x$ to the neighboring site $x+\hat\mu$ in
the $\mu$-direction. In the classical continuum limit the standard action
with $\beta=1/g^2$ reduces to the continuum action of eq.~\eqref{Scont}. 
According to Symanzik's effective theory, 
the standard action is expected to have cutoff effects of 
$\cO(a^2)$. Recently, it has been observed that large logarithmic 
corrections, which also result from Symanzik's theory \cite{Bal09}, can mimic 
the apparent $\cO(a)$ behavior that was observed in numerical simulations 
\cite{Has02,Kne05}.

Recently, we have performed detailed investigations of so-called topological
lattice actions \cite{Bie10}, which are invariant against small deformations of 
the lattice field. Here we investigate a topological action that constrains the 
angle between neighboring spins by a maximal value $\delta$. 
The topological action vanishes, i.e.\ $S_{\mathrm{top}}[\vec e] = 0$, if the
constraint is satisfied for all nearest-neighbor spin pairs, 
i.e.\ $\vec e_x \cdot \vec e_{x+\hat\mu} > \cos\delta$, and is infinite 
otherwise. This action has already been considered 
earlier in 
\cite{Pat92,Pat93,Aiz94,Has96,Pat02}, 
without emphasizing its
topological features. Since the topological action vanishes for all allowed
configurations, it has no meaningful classical continuum limit and cannot be
treated in perturbation theory. Still, when one sends $\delta \rightarrow 0$,
one approaches the correct quantum continuum limit \cite{Has96,Bie10}, which 
demonstrates that universality does not rely on the classical continuum limit. 
Interestingly, for the topological action the sign of the cutoff effects of 
some observables is opposite to the one resulting from the standard action.

In \cite{Boe11} we have used this observation to construct an optimized action
with extremely small cutoff effects. The resulting constrained action combines 
the standard and the topological action such that
\begin{equation}
  \label{Scons}
  S_{\mathrm{con}}[\vec e] = \sum_{x,\mu} s(\vec e_x,\vec e_{x+\hat\mu}),
\end{equation}
with
\begin{equation}
  s(\vec e_x,\vec e_{x+\hat\mu}) = 
  \beta 
  (1 - \vec e_x \cdot \vec e_{x+\hat\mu}) \qquad \mathrm{for} \quad 
  \vec e_x \cdot \vec e_{x+\hat\mu} > \cos\delta,
\end{equation}
and $s(\vec e_x,\vec e_{x+\hat\mu}) = \infty$ otherwise. For $\delta = \pi$ the
constrained action reduces to the standard action, while for $\beta = 0$ it
turns into the topological action. By optimizing the constraint angle to the
value $\cos\delta = - 0.345$, we have been able to reduce the cutoff effects 
below the per mille level in an investigation of lattice $\theta$-vacua in the 
2-d $\mrO(3)$ model \cite{Boe11}. When $\delta$ takes the optimized value, the 
constrained action $S_{\mathrm{con}}[\vec e]$ turns into the optimized 
constrained action $S_{\mathrm{oca}}[\vec e]$. 
In this paper, we construct such actions also for $N \neq 3$.

Finally, we consider the following quadratic plus quartic mixed action
\begin{equation}
  \label{Smix}
  S_{\mathrm{mix}}[\vec e] = \sum_{x,\mu} \left[ 
    \beta
    (1 - \vec e_x \cdot \vec e_{x+\hat\mu}) +
    \gamma (1 - \vec e_x \cdot \vec e_{x+\hat\mu})^2\right].
\end{equation}
For given $\beta$, we will adjust $\gamma$ to reduce the cutoff effects, 
thus  optimizing the action. When $\gamma \gtrsim \beta^2$ the mixed action 
can again not  be treated in perturbation theory. 
When $\gamma$ takes the optimized value, 
the mixed action $S_{\mathrm{mix}}[\vec e]$ turns into the optimized mixed 
action $S_{\mathrm{oma}}[\vec e]$. 

\begin{figure}[htb]
  \centering
  \includegraphics[width=0.4\textwidth,angle=270]{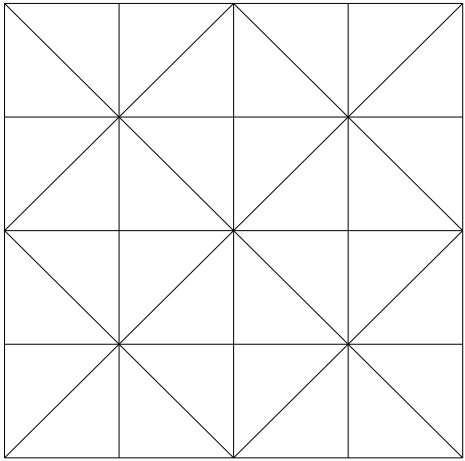}
  \caption{\it Decomposition of the square lattice into triangles 
    $t_{xyz}$. The topological charge density $A_{xyz}/4 \pi$ of the 2-d 
    lattice $\mrO(3)$ model is given by the oriented area $A_{xyz}$ of 
    the spherical triangle defined by the three spins $\vec e_x$, 
    $\vec e_y$, and $\vec e_z$ at the corners of $t_{xyz}$.}
  \label{triangles}
\end{figure}

Berg and L\"uscher have introduced a geometric definition of the lattice 
topological charge \cite{Ber81}. In this definition, each lattice plaquette
is divided into two triangles, as illustrated in figure~\ref{triangles}. 
The spins $\vec e_x$, $\vec e_y$, and $\vec e_z$ at the three corners of a 
lattice triangle $t_{xyz}$ define the corners of a spherical triangle on $S^2$. 
The oriented area $A_{xyz}$ of the spherical triangle is given by
\begin{eqnarray}
  \label{area}
  &&A_{xyz} = 2 \varphi \in (- 2 \pi,2 \pi], \quad 
  X + i Y = r \exp(i \varphi), \nonumber \\
  &&X = 1 + 
  \vec e_x \cdot \vec e_y + \vec e_y \cdot \vec e_z + \vec e_z \cdot \vec e_x,
  \quad Y = \vec e_x \cdot (\vec e_y \times \vec e_z).
\end{eqnarray}
The lattice topological charge is defined as the sum of the oriented areas 
$A_{xyz}$ over all triangles $t_{xyz}$, normalized by the total area $4 \pi$ of 
the sphere $S^2$, i.e.
\begin{equation}
  Q[\vec e] = \frac{1}{4 \pi} \sum_{t_{xyz}} A_{xyz} \in \Z.
\end{equation}
The decomposition of the square lattice into triangles illustrated in 
figure~\ref{triangles} is invariant under 90 degrees rotations and 
under translations by an even number of lattice spacings. 
When a configuration is shifted by just
one lattice spacing, the lattice plaquettes are divided into two triangles in
a different manner than before, and thus the topological charge may change.
However, this does not happen for sufficiently smooth configurations. In
particular, one can show that a nearest-neighbor constraint angle 
$\delta < \pi/2$ (as used in the action $S_{\mathrm{con}}$) leads to a completely
translation invariant topological charge. In that case, just as in the 
continuum, different topological sectors are separated by infinite-action 
barriers.

\section{Numerical Study of Cutoff Effects in the 2-d 
  \boldmath $\mrO(3)$ Model}
\label{NumStudy}

In this section we investigate the cutoff effects of a variety of physical
quantities. In particular, we address the question to what extent an action
that was optimized to reproduce the continuum limit of a particular quantity
automatically improves the scaling behavior of other observables.

\subsection{The step scaling function at \boldmath $\theta = 0$}
\label{StepScaling} 

The dimensionless physical quantity 
\begin{equation}
  u = L m(L) = L/\xi(L),
\end{equation}
is defined as the ratio of the spatial size $L$ and the finite-volume 
correlation length $\xi(L) = 1/m(L)$. Based on this, one defines
the step scaling function (with scale factor $s$) \cite{Lue91}
\begin{equation}
  \sigma(s,u) = s L m(s L),
\end{equation}
which is also known analytically \cite{Bal04}. By measuring the mass gaps
$m(L)$ and $m(sL)$ in a Monte Carlo simulation on a lattice with $L/a$ points
in the periodic spatial direction, one obtains the lattice step scaling function
$\Sigma(s,u,a/L) = s L m(sL,a/L)$.

\begin{figure}[t]
  \centering
  \includegraphics[width=0.8\textwidth,angle=0]{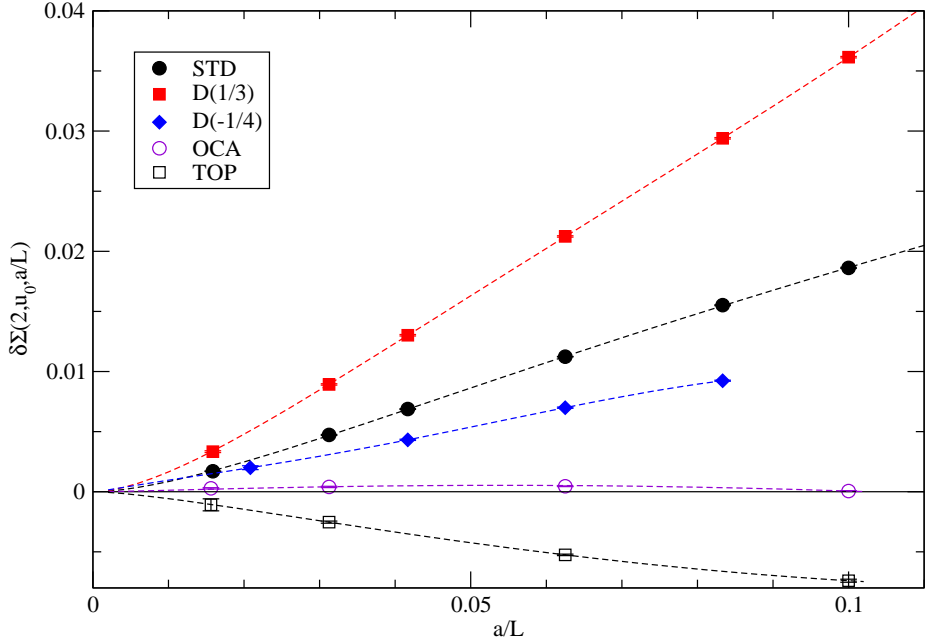}
  \caption{\it Cutoff dependence of the step scaling function 
    $\Sigma(2,u_0,a/L)$ for five different lattice 
    actions: the standard action, the actions $D(1/3)$ and $D(-1/4)$ with 
    additional diagonal neighbor couplings, the topological lattice action 
    of \cite{Bie10}, and the optimized constrained action with 
    $\cos\delta = - 0.345$. The lines for the standard action
    are fits based on eq.~\eqref{fit} . 
    The horizontal line represents the analytic continuum result of 
    \cite{Bal04}.}
  \label{step2}
\end{figure}

Figure~\ref{step2} compares the cutoff effects of $\Sigma(2,u_0,a/L)$ 
at $u_0 = 1.0595$ for five different lattice actions. While the standard action 
and the actions $D(1/3)$ and $D(-1/4)$ \cite{Bal10a} (which contain additional 
diagonal neighbor couplings ), approach the continuum limit from above, the 
topological action approaches it from below. By optimizing the constraint angle 
$\delta$ such that $\Sigma(2,u_0,a/L=1/10)$ assumes its continuum value
$\sigma(2,u_0) = 1.26121035$, 
one obtains $\cos\delta = - 0.345$.
The resulting optimized constrained action has extremely small cutoff effects 
in the per mille range also at other (not too coarse) values of the lattice 
spacing. Based on the analytic results of \cite{Bal10a} the lattice data have 
been fitted to the expression
\begin{equation}
  \label{fit}
  \Sigma(2,u,a/L) = \sigma(2,u) + \frac{a^2}{L^2} 
  \left[B \log^3(L/a) + C \log^2(L/a) + \dots\right],
\end{equation}
which gives good agreement with the analytically known continuum result. The
above analytic form is justified theoretically for the standard action as well
as for the actions $D(1/3)$ and $D(-1/4)$. Although, strictly speaking, it may
no longer be justified to use the expression of eq.~\eqref{fit} to 
fit the topological action data, one again obtains excellent agreement with the
continuum result. Interestingly, in the range of correlation lengths considered 
here, the lattice artifacts of the topological action are smaller than those of 
the standard action. In fact, for the standard action at $L/a = 64$ the 
sub-leading term proportional to $\log^2(L/a)$ is still larger than the 
leading term proportional to $\log^3(L/a)$, while this is not the case for the
topological action \cite{Bie10}. Based on the fitted values of $B$ and $C$,
the lattice artifacts of the standard action are smaller than the ones of the 
topological action only for correlation lengths larger than about 
$5 \times 10^4 a$.
Without re-adjusting $\delta$, using the optimized constrained action (which 
was optimized for $u_0 = 1.0595$ at $L/a = 10$), the cutoff effects of the step
scaling function are extremely small also for other values of $u$. Of course,
one could also re-adjust $\delta$ for each value of $L/a$ (always optimizing
at $u_0$). Interestingly, the optimal value of $\delta$ is rather
insensitive to $L/a$ (as long as $L/a$ is not too small).

\begin{figure}[t]
  \centering
  \includegraphics[width=0.8\textwidth,angle=0]{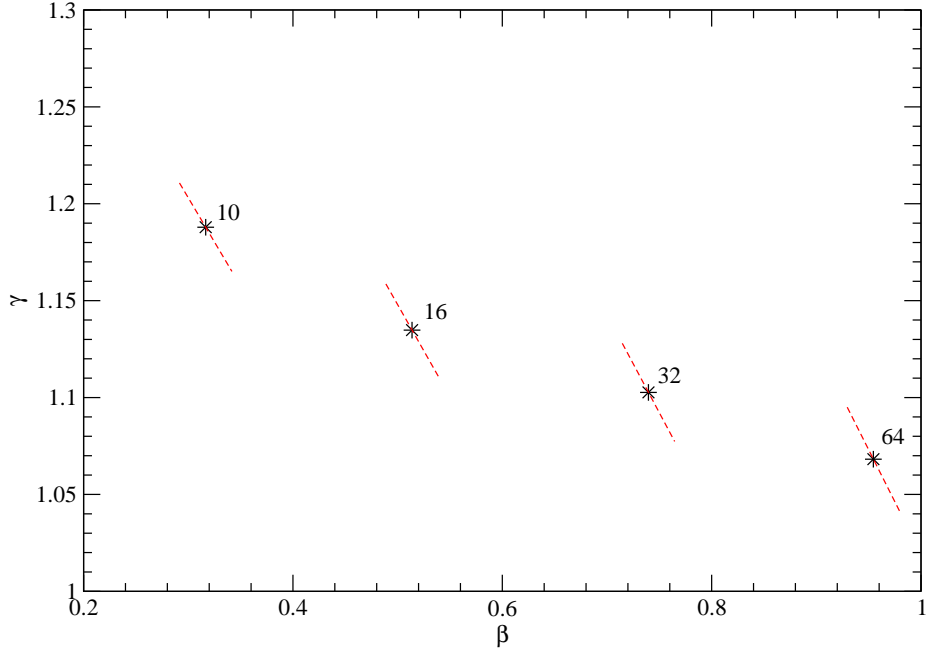}
  \caption{\it Optimized parameters $\beta$ and $\gamma$ 
    for the mixed action such that 
    $\Sigma(2,u_0,a/L) = \sigma(2,u_0) = 1.26121035$.
    The dashed lines represent the tangents to lines of constant 
    $\Sigma(2,u_0,a/L)$, which are obtained by measuring the derivatives 
    $\p u/\p \beta$ and $\p u/\p \gamma$.}
  \label{opt_bg}
\end{figure}

Let us also pursue this more elaborate alternative optimization strategy, 
however, now applied to the mixed action. For each value of $\beta$, the 
parameter $\gamma$ of the mixed action has been optimized to reproduce the 
analytically known continuum limit of the step 2 scaling function, 
i.e.\ $\Sigma(2,u_0,a/L) = \sigma(2,u_0) = 1.26121035$. 
The optimal values of $\beta$ and $\gamma$ are shown in 
figure~\ref{opt_bg} for various lattice sizes $L/a$. 
They are also listed in table~\ref{optimal}.

\begin{table}
  \centering
  \begin{tabular}{|c|c|c|c|c|c|}
    \hline
    $L/a$ & $\beta$ & $\gamma$ & $u$ & $\Sigma(2,u,a/L)$ & 
    $\Sigma(2,u_0,a/L) - \sigma(2,u_0)$ \\
    \hline
    \hline
    10 & 0.31664 & 1.18790 & 1.059500(6)\phantom{0} & 1.261218(12) & 
    $\phantom{-}0.000008(16)$ \\
    \hline
    16 & 0.51378 & 1.13480 & 1.059506(9)\phantom{0} & 1.261204(12) & 
    $-0.000016(19)$ \\
    \hline
    32 & 0.73962 & 1.10276 & 1.059505(9)\phantom{0} & 1.261167(17) & 
    $-0.000011(22)$ \\
    \hline 
    64 & 0.95450 & 1.06820 & 1.059501(15)& 1.261205(36) & $-0.000007(44)$ \\
    \hline
  \end{tabular}
  \caption{\it Optimized parameters $\beta$ and $\gamma$ for 
    the mixed action such that 
    $\Sigma(2,u_0,a/L) = \sigma(2,u_0) = 1.26121035$. 
        The resulting value of $u$ is very close to the desired $u_0 = 1.0595$, 
    and $\Sigma(2,u,a/L)$ deviates very little from the continuum value 
    $\sigma(2,u_0)$.
    The last column is the deviation of $\Sigma(2,u_0,a/L)$ (obtained by 
    extrapolation from $u$ to $u_0$) from the analytic result.}
  \label{optimal}
\end{table}

\begin{figure}[t]
  \centering
  \includegraphics[width=0.8\textwidth,angle=0]{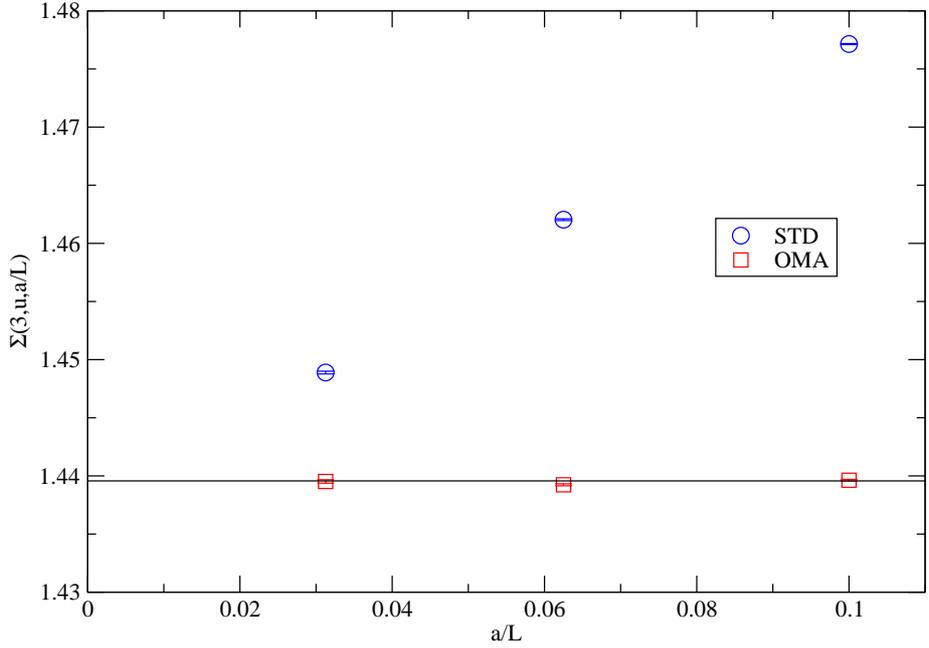}
  \caption{\it Cutoff dependence of the step 3 scaling function 
    $\Sigma(3,u_0,a/L)$ for the standard action and the optimized 
    mixed action.}
  \label{step3}
\end{figure}

Using the couplings of table~\ref{optimal}, figure~\ref{step3} shows the 
resulting cutoff effects of the step 3 scaling function $\Sigma(3,u_0,a/L)$ 
compared to the analytically known continuum result 
$\sigma(3,u_0) = 1.439574$. 
Without any further adjustable parameters, the mixed action optimized in this
way automatically leads to a drastic reduction of the cutoff effects of
$\Sigma(3,u_0,a/L)$.

\subsection{The step scaling function at \boldmath $\theta \neq 0$}
\label{StepTheta}

While analytic results from the exact S-matrix exist only for $\theta = 0$ and
$\theta = \pi$, the step scaling function is well-defined for all values of 
the vacuum angle $\theta$. In \cite{Boe11} the step 2 scaling function has been
determined at $\theta = 0, \pi/2$, and $\pi$. In this way, the 
conjectured exact S-matrix has been indirectly verified with per mille level
accuracy. It also has been shown that the step scaling function at 
$\theta = \pi/2$ differs from the one at $\theta = 0$ and $\pi$. This
shows that $\theta$ does not get renormalized non-perturbatively and it 
indicates that all values of $\theta \in [0,\pi]$ are associated with distinct 
theories in the continuum limit. 
In view of the potentially devastating cutoff
effects caused by dislocations, this is a non-trivial and somewhat 
unexpected result. 
Note, however, that a non-trivial $\theta$-dependent
spectrum is predicted by form-factor perturbation
theory \cite{Muss03}. The spectrum was investigated by
lattice Monte Carlo methods \cite{Alles07} using an imaginary
$\theta$ and analytic continuation.

\begin{figure}[t]
  \centering
  \includegraphics[width=0.8\textwidth,angle=0]{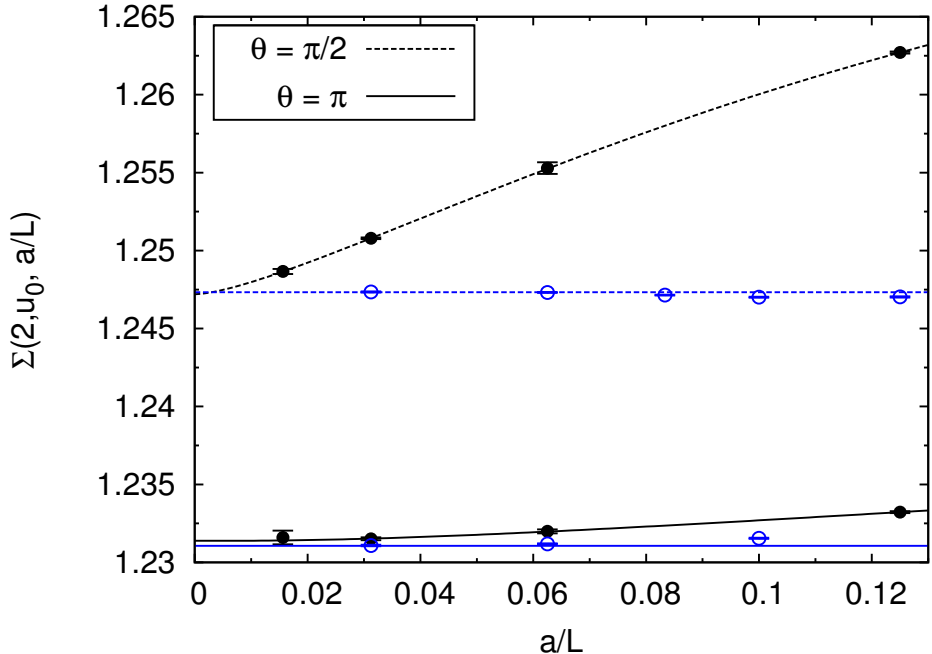}
  \caption{\it Cutoff dependence of the step scaling function 
    $\Sigma(\theta,2,u_0,a/L)$ (with $L m(\theta,L) = u_0 = 1.0595$) for the 
    standard and for the optimized constrained action with 
    $\cos\delta = - 0.345$, at $\theta = \pi/2$ and $\theta = \pi$. 
    The lines for the standard action are fits based on eq.~\eqref{fit}.
    The horizontal lines represent the analytic result of \cite{Bal11} 
    at $\theta = \pi$, 
    and the fitted continuum value for $\theta = \pi/2$.}
  \label{steptheta}
\end{figure}

Figure~\ref{steptheta}, which is taken from \cite{Boe11}, shows that 
without any further adjustments of $\delta$, the cutoff effects of the 
optimized constrained action are automatically drastically reduced also at 
$\theta = \pi/2$ and $\pi$. The analytic result is known \cite{Bal11} 
for $\theta=\pi$ to be $\sigma(\pi,2,u_0)=1.231064$ for $u_0=1.0595$.

\begin{figure}[t]
  \centering
  \includegraphics[width=0.8\textwidth,angle=0]{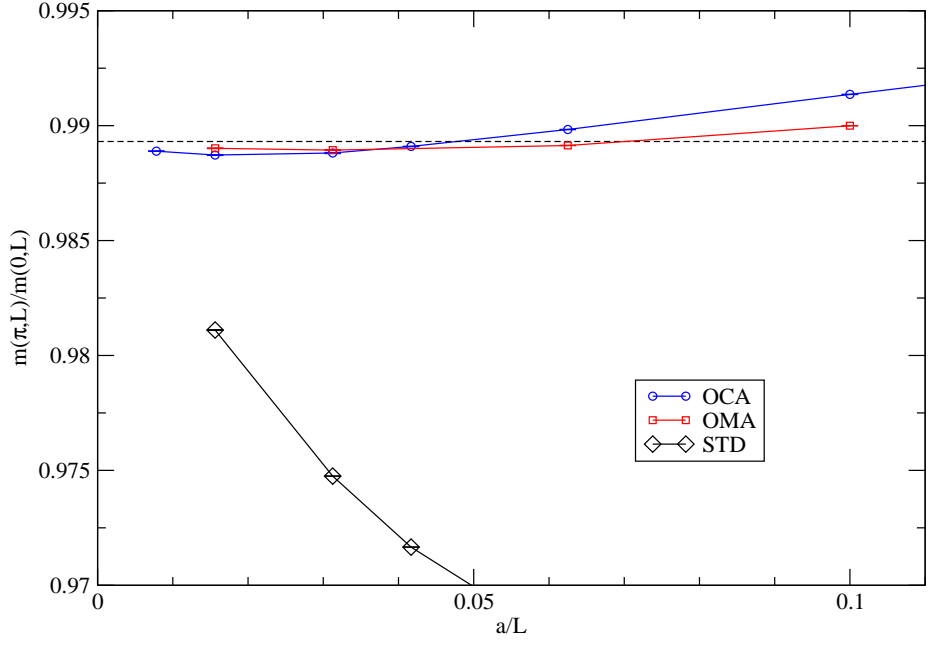}
  \caption{\it Cutoff dependence of the mass gap ratio 
    $m(\theta = \pi,L)/m(0,L)$ at fixed $L m(0,L) = u_0 = 1.0595$
    for the standard, constrained (with $\cos \delta=-0.345$), 
    and optimized mixed action. 
    The horizontal line is the analytic result.}
  \label{mr_pi}
\end{figure}

We have also considered $L m(\theta,L)$ at fixed $L m(0,L) = u_0 = 1.0595$, 
which is yet another physical quantity. 
Figure~\ref{mr_pi} illustrates the cutoff effects of the mass gap at 
$\theta = \pi$ for the optimized constrained action 
(with $cos\delta=-0.345$), the optimized mixed action and the 
standard action.
Although the cutoff effects are only in the per mille range for the
optimized actions, the approach to the continuum limit is non-uniform 
since the lattice results undershoot the exact continuum value before 
they ultimately approach it from below.

\begin{figure}[t]
  \centering
  \includegraphics[width=0.8\textwidth,angle=0]{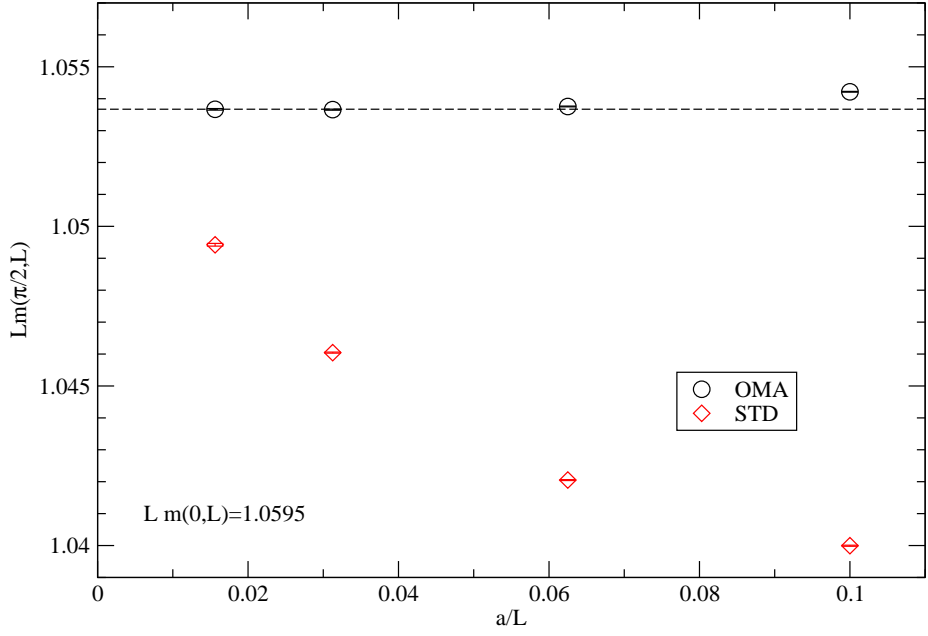}
  \caption{\it Cutoff dependence of $L m(\theta,L)$ at fixed 
    $L m(0,L) = u_0 = 1.0595$, for the standard and for the optimized 
    mixed action at $\theta = \pi/2$. 
    The horizontal line represents the fitted continuum value.}
  \label{u_pi2}
\end{figure}

We have also measured $L m(\pi/2,L)$ at fixed $L m(0,L) = u_0 = 1.0595$
using the optimized mixed action with the couplings listed in 
table~\ref{optimal}. 
As shown in figure~\ref{u_pi2}, the cutoff effects are 
again drastically smaller than for the standard action.

\subsection{The mass of the isotensor state}
\label{Isotensor}

\begin{figure}[t]
  \centering
  \includegraphics[width=0.8\textwidth,angle=0]{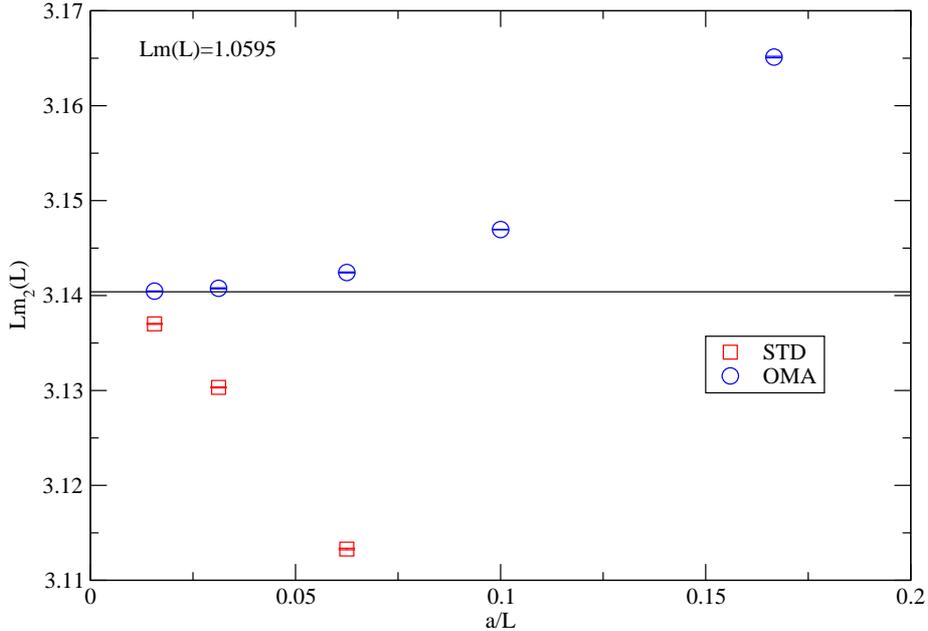}
  \caption{\it Cutoff effects of the mass of the isotensor state
    for the standard and optimized mixed action (with the parameters
    listed in table~\ref{optimal}).}
  \label{I2_O3}
\end{figure}

Besides the state in the isovector channel, 
the exact S-matrix also provides
analytic results for the finite-volume mass $m_2(L)$ 
in the isotensor channel ($I=2$). 
Its approach to the continuum limit is shown in figure~\ref{I2_O3} 
for both the optimized mixed and the standard action. 
Again, the optimized mixed action reaches the
continuum limit much faster than the standard action.

\subsection{The renormalized coupling \boldmath $\gR$}
\label{Renormalized}

\begin{figure}[t]
  \centering
  \includegraphics[width=0.8\textwidth,angle=0]{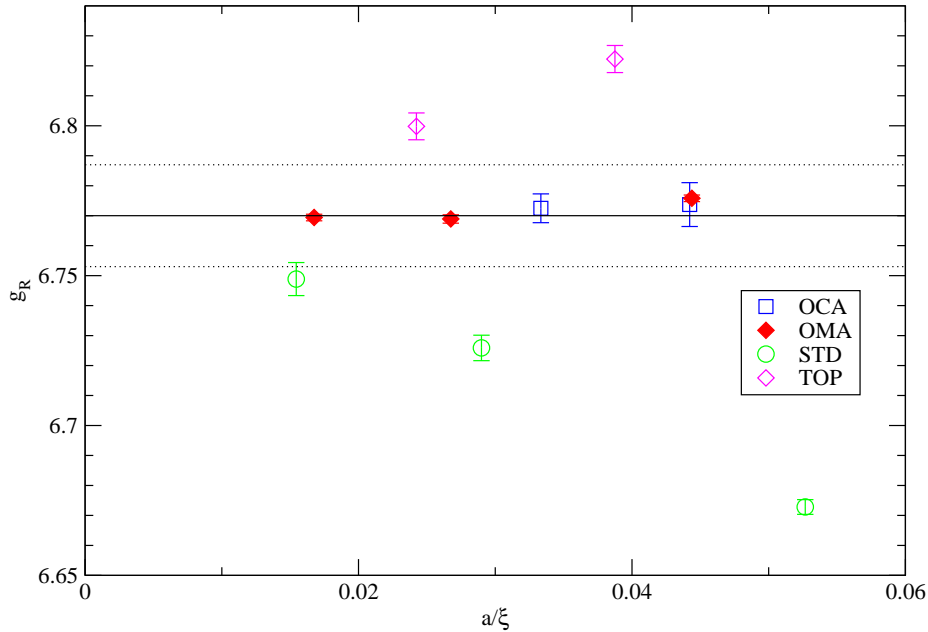}
  \caption{\it Cutoff effects of the renormalized coupling $\gR$ for four
    different lattice actions. The thick horizontal line is the result 
    $\gR = 6.770(17)$ obtained from the exact S-matrix (whose estimated
    theoretical error corresponds to the dotted horizontal lines) 
    \cite{Bal00}.}
  \label{gR}
\end{figure}

The renormalized coupling 
defined in terms of a truncated 4-point function
at zero momentum (see eq.~\eqref{gRdef}), has been calculated from the 
exact S-matrix in \cite{Bal00} with the result $\gR=6.770(17)$.
As illustrated in figure~\ref{gR}, without any further tuning 
of the optimized actions, the continuum limit of $\gR$ is approached much more 
quickly with the optimized constrained and the optimized mixed action than 
with the standard or the topological action. 
The results from the optimized actions suggest that the theoretical 
error given in \cite{Bal00} has been overestimated.
Based on the last two points with the optimized mixed action
here we obtain the value $\gR=6.769(2)$. 

\section{Numerical Study of Cutoff Effects in the 2-d \boldmath 
  $\mrO(4)$  and $\mrO(8)$ Models}
\label{Num_O4_O8}

In order to investigate whether highly optimized local actions can also be
constructed successfully for larger values of $N$, in this section we study
the 2-d $\mrO(4)$ and the 2-d $\mrO(8)$ model. 
In these cases we limit ourselves to the optimized constrained action 
compared to the standard action.

\subsection{The step scaling function of the 2-d \boldmath $\mrO(4)$ model}
\label{SSF_O4}
\begin{figure}[t]
  \centering
  \includegraphics[width=0.8\textwidth,angle=0]{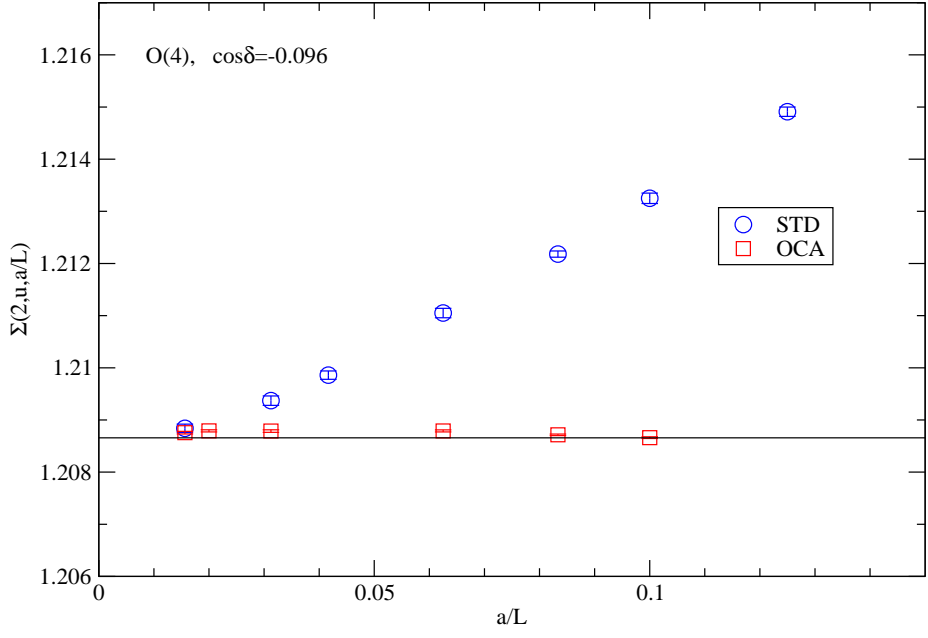}
  \caption{\it Cutoff effects of the step-scaling function in the 2-d 
    $\mrO(4)$ model. The horizontal line indicates the exact result in 
    the continuum limit, $\sigma(2,1.0)=1.208658$. 
    The optimized constrained action has cutoff 
    effects in the fraction of a per mille range.}
  \label{step2_O4}
\end{figure}

Figure~\ref{step2_O4} compares the cutoff effects of the step 2 scaling 
function for the standard action with those of the optimized constrained 
action. The constraint angle $\delta$ has been optimized by demanding
$\Sigma(2,u_0,a/L=1/10) = \sigma(2,u_0)$, in this case for $u_0 = 1$, which
yields $\cos\delta = - 0.096$. Again, the cutoff effects are drastically
reduced in comparison to the standard action. Still, there are remaining 
tiny cutoff effects in the per mille range. The cutoff effect is not
monotonic, after a small increase it starts to diminish only on an 
$L/a = 32$ lattice.
The data for the standard action are taken from \cite{Bal09}.

\subsection{The step scaling function of the 2-d \boldmath  $\mrO(8)$ model}
\label{SSF_O8}

\begin{figure}[t]
  \centering
  \includegraphics[width=0.8\textwidth,angle=0]{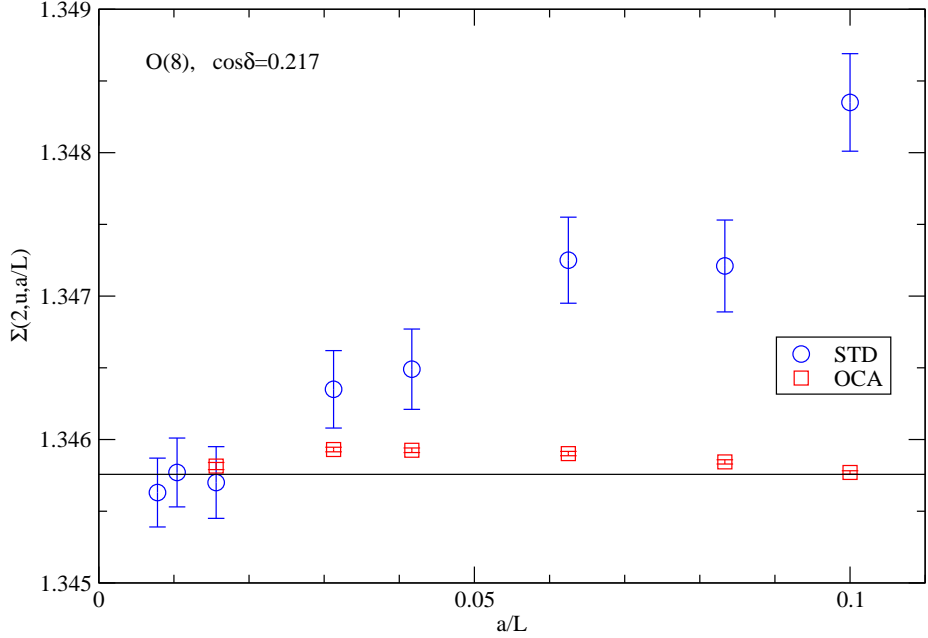}
  \caption{\it Cutoff effects of the step-scaling function in the 2-d 
    $\mrO(8)$ model. The horizontal line indicates the exact result in 
    the continuum limit, $\sigma(2,1.0595)=1.345757$. 
    The optimized constrained action has cutoff 
    effects in the fraction of a per mille range.}
  \label{step2_O8}
\end{figure}

Figure~\ref{step2_O8} shows the analogous results for the $\mrO(8)$ case.
In this case, the constraint is $\cos\delta = 0.217$, again for $u_0=1.0595$.
The data for the standard action are from \cite{Kne05}.

\section{Analytic Study of Cutoff Effects in the 2-d \boldmath 
  $\mrO(N)$ Model at $N = \infty$}
\label{N_infinity}

The step scaling function for the $\mrO(N)$ non-linear sigma model 
in the limit $N\to\infty$ has been studied in \cite{Lue82c,CaPe,Kne05}.
Here we generalize these calculations for the case of the 
constrained action and of the mixed action.
Note that here we use lattice units, ``$a=1$''.

\subsection{Constrained action in the large \boldmath $N$ limit }
\label{N_inf_CA}

The action and the partition function in the constrained case
are given by
\begin{equation}
  S[\evec] = \frac{N}{2f}\sum_{x,\mu} (\p_\mu \evec_x)^2 \,,
\end{equation}

\begin{equation} \label{Z1} 
  Z = \int_{\evec} \exp\Big\{ - S[\evec] \Big\}
  \prod_x \delta( \evec_x^2-1) \prod_{x,\mu} \Theta( \epsilon - (\p_\mu
  \evec_x)^2) \,.
\end{equation}
Here $\Theta$ is the step function, $\p_\mu$ denotes the 
forward lattice derivative, and 
$\int_{\evec}$ denotes $\int\prod_x {\mrd\evec_x}$.
As usual, the $N\to\infty$ limit is taken with $\beta/N=1/f$ fixed.

For the standard action the distribution of $(\p_\mu \evec_x)^2$ in the
large $N$ limit approaches a $\delta$-function at 
$(\p_\mu \evec_x)^2=f/d$ (where $d$ is the number of Euclidean 
dimensions),
as can be easily obtained from perturbation theory. 
On the other hand, for the topological
action (with only the $\Theta$-constraint, 
i.e.\ with $f=\infty$) it goes also to a $\delta$-function 
at $(\p_\mu \evec_x)^2=\epsilon$.  One expects
that for $\epsilon > f/d$ the constraint is irrelevant (it is above the narrow
peak) and $f$ alone determines the physics in the large $N$ limit. 
On the contrary, at $\epsilon < f/d$ the constraint dominates  the physics.
It is also clear from these considerations that at $N=\infty$ the 
$\Theta(\epsilon - (\p_\mu \evec_x)^2)$ factors can be replaced by a 
strict constraint $\delta(\epsilon - (\p_\mu \evec_x)^2)$.

Introducing the auxiliary variables $\alpha_x$, $\eta_{x\mu}$ one gets
\begin{equation} \label{Z1a} 
  Z = \int_{\evec, \alpha, \eta} \exp\Big\{ -
  S[\evec] + i (\alpha, \evec^2-1) 
  + i (\eta, \epsilon-(\p \evec)^2)
  \Big\} \frac{1}{\eta-i0}  \,.
\end{equation}
Here the auxiliary variables are integrated over the real axis.

Note that $\epsilon= 2 (1-\cos\delta)$ where $\delta$ is the maximal allowed
angle between the nearest neighbors. The allowed range is 
$0 < \epsilon \le 4$.  For $\epsilon \ge 4$ one recovers the standard 
(unconstrained) action.

Rescaling $\alpha$ and $\eta$ as
\begin{equation}
  \alpha_x \to \frac12 N \alpha_x \,,\qquad 
  \eta_{x\mu} \to \frac12 N\eta_{x\mu} \,,
\end{equation}
and then shifting the integration contour\footnote{%
  Due to eq.~\eqref{Z1a} the integration line of $\eta_{x\mu}$ 
  can be shifted only downwards, i.e. one should have $v_\mu \ge 0$.}
\begin{equation} \label{shift}
  \alpha_x \to i h + \frac{\alpha_x}{\sqrt{N}} \,,\qquad 
  \eta_{x\mu} \to -i v_\mu + \frac{\eta_{x\mu}}{\sqrt{N}}  \,,
\end{equation}
one obtains a form suitable for the $1/N$ expansion: 
\begin{multline} \label{Aeff}
  S_{\mathrm{eff}} = \frac{N}{2} \left[ \sum_{x,\mu} \left(
      \frac{1}{f}+v_\mu\right)(\p_\mu \evec_x)^2 
    + h \sum_x \evec_x^2 - V h -
    \epsilon V\sum_\mu v_\mu
  \right] \\
  -i \frac{\sqrt{N}}{2} \Big[ \sum_x \alpha_x (\evec_x^2 -1) + \sum_{x,\mu}
  \eta_{x\mu} (\epsilon - (\p_\mu \evec_x)^2) \Big]
  +\sum_{x,\mu}\ln\left(v_\mu + i\frac{\eta_{x\mu}}{\sqrt{N}}\right) \,,
\end{multline}

Note that in eq.~\eqref{shift} we have allowed for different values
of $v_\mu$ in different directions.
In symmetric volumes $V=L^d$ they can be replaced by a single $v$.
For the step scaling function one should take the strip geometry,
$\infty \times L^{d-1}$.
Here one should expect different $v_\mu$'s in the time- and spatial 
directions.\footnote{Note that one could avoid this complication
by modifying the constraint to an ``isotropic'' one,
$\Theta\left( \epsilon - \overline{(\p \evec_x)^2}\right)$,
meaning that the average of $(\p_\mu \evec_x)^2$ over the $2d$
nearest neighbors should not exceed $\epsilon$.}

After integrating out the spin fields we get\footnote{In the 
leading order in $1/N$ we will not need
the terms depending on the auxiliary fields $\alpha_x$, $\eta_{x\mu}$.
Hence they are not written out explicitly here.}
\begin{equation} \label{Aeffbar}
  \overline{S}_\mathrm{eff} = \frac{N}{2}\left[ \mathrm{tr}\ln S -
    V h - V \epsilon \sum_\mu v_\mu
  \right] + \cO\left(\sqrt{N}\right) \,,
\end{equation}
with
\begin{align}
  & S_{xy} = h \delta_{xy} +\sum_\mu w_\mu
  \left[2\delta_{xy}-\delta_{y,x+\hat{\mu}}-\delta_{y,x-\hat{\mu}}\right] \,,
  \label{Sxy} \\
  & w_\mu = \frac{1}{f}+v_\mu \label{wmu} \,.
\end{align}

For the inverse we have
\begin{equation} \label{Sinv}
  S^{-1}_{xy} = \frac{1}{V}\sum_p\frac{\mre^{ip(x-y)}}{D(p)} \,,
\end{equation}
\begin{equation} \label{Dp}
  D(p)=\sum_\mu w_\mu K_\mu(p) + h \,,
\end{equation}
where the sum is over 
$p_\mu=2\pi n_\mu/L_\mu\,,\,\,\,n_\mu=0,1,\ldots,L_\mu-1$ 
and where $K_\mu(p)=2(1-\cos p_\mu)$.

Note that the $1/(\eta-i0)$ denominator (i.e. the logarithmic term
in eq.~\eqref{Aeff}) does not enter in the $N=\infty$ limit, 
i.e. the $\Theta$-constraint is effectively replaced by a
$\delta$-constraint, as discussed above.

Eq.~\eqref{Aeffbar} in leading order yields the gap equations 
\begin{align} \label{gap01}
  & \frac{1}{V} \sum_p \frac{1}{D(p)} = 1 \,,
  \\
  \label{gap02}
  & \frac{1}{V} \sum_p \frac{K_\mu(p)}{D(p)} = \epsilon \,.
\end{align}
Consistency of these equations gives the relation
\begin{equation} \label{heps}
   h +\epsilon\sum_\mu w_\mu = 1\,.
\end{equation}

The finite volume mass gap $m(L)$ in the strip geometry 
$\infty\times L^{d-1}$ in leading order $1/N$ is given by
\begin{equation} \label{m0m}
  m_0=2\sinh \frac{m(L)}{2} \,,
\end{equation}
where
\begin{equation} \label{m0hw0}
  m_0^2 = \frac{h}{w_0} \,.
\end{equation}
Defining
\begin{equation} \label{umL}
  u = m(L)L \,,
\end{equation}
one has
\begin{equation} \label{m0u}
  m_0 = m_0(u,L) = 2 \sinh \frac{u}{2L} =
  \frac{u}{L}\left( 1 + \frac{u^2}{24 L^2}+\ldots \right) \,.
\end{equation}
The continuum limit is approached by $L\to\infty$ keeping $u$ fixed.

\subsubsection{The isotropic case}
\label{N_inf_iso}

In a symmetric volume $L^d$ (or with the modified ``isotropic action'')
one has $v_\mu=v$ and the gap equations 
\eqref{gap01}, \eqref{gap02} yield
\begin{equation} \label{gap_is} 
  \frac{1}{V} \sum_p \frac{1}{K(p) + m_0^2} =
  \frac{1}{m_0^2 + \epsilon d} \,,
\end{equation}
as well as the relation
\begin{equation}
  \left(v+\frac{1}{f}\right)\epsilon d + h = 1 \,.
\end{equation}
Since one should have $h$ and $v$ non-negative, this relation has a solution
only for $ \epsilon \le f/d$.  As discussed above, for larger $\epsilon$ the
kinetic term is the relevant one and the constraint can be omitted.
As expected, the gap equation \eqref{gap_is} has a solution only for
$\epsilon < 4$ since for the standard action $0 \le K(p) \le 4$.

For the standard (unconstrained) action the gap equation reads
\begin{equation} \label{gap_st} 
  \frac{1}{V} \sum_p \frac{1}{K(p) + m_0^2} =
  \frac{1}{f} \,.
\end{equation}
In the infinite volume eq.~\eqref{gap_is} gives
\begin{equation} \label{gap_inf} 
  \int_{-\pi}^\pi \frac{\mathrm{d}^dp}{(2\pi)^d} 
  \frac{1}{K(p) + m_0^2} =
  \frac{1}{m_0^2 + \epsilon d} \,.
\end{equation}

From now on we consider the $d=2$ case. In this case the integral on the
l.h.s. of eq.~\eqref{gap_inf}  diverges logarithmically like 
$1/(4\pi) \log(a^2 m_0^2)$, where we restored the lattice spacing.  
For the standard action this gives the well known result
\begin{equation} \label{am} 
  m \approx \frac{\mathrm{const}}{a} \mre^{-2\pi/f} \,.
\end{equation}

Since $m_0^2 \propto \exp(-2\pi/\epsilon)$ in the continuum limit
$\epsilon \to 0$ the $m_0^2$ term can be neglected on the r.h.s. 
of eqs.~\eqref{gap_is}, \eqref{gap_inf}, and in the continuum limit 
one recovers the universal result.
The cutoff effects will be, however, slightly different.  
To determine them for the step scaling function one
can solve the gap equation \eqref{gap_is} numerically in a
finite volume for different lattice spacings (i.e.\ different $L/a$).
It turns out that the leading $\cO(a^2)$ cutoff effects can be calculated
analytically.

\subsubsection{The lattice artifacts at \boldmath $N=\infty$ 
for the step scaling function}
\label{N_inf_CA_art}

Consider the $d=2$ case in the strip geometry $\infty\times L$ 
and introduce
\begin{equation} \label{rhodef}
  \rho = \frac{w_1}{w_0}\,,
\end{equation}
\begin{equation} \label{omegadef}
  \omega = \rho K_1(p) + m_0^2 \,.
\end{equation}

Writing eqs.~\eqref{gap01}, \eqref{gap02} for $L_t=\infty$ 
after integrating out $p_0$ one obtains
a set of two equations
\begin{align} \label{gap1}
  \frac{1}{L}\sum_p \frac{1}{\sqrt{\omega(\omega+4)}} & = w_0 \,,
   \\ \label{gap2}
  \frac{1}{L}\sum_p \frac{K_1(p)}{\sqrt{\omega(\omega+4)}} & =
   \epsilon w_0 \,,
\end{align}
where from eq.~\eqref{heps} one has $w_0 = 1/ [ m_0^2 + (\rho+1)\epsilon] $. 

We are interested in the approach of the 
step scaling function $u'(u,L)=m(2L)2L$ for fixed $u(L)=m(L)L$
to the continuum limit $a/L\to0$. Analytic results are obtained
using the results of Caracciolo and Pelissetto \cite{CaPe};
technical details are deferred to appendix~\ref{AppB1}.
The result is of the form
\begin{equation} \label{uprime}
  u'=u^{\prime}_{\infty}+\frac{1}{L^2}\nu(u,z) + \ldots \,,
\end{equation}
where the continuum limit $u^{\prime}_{\infty}=\sigma(2,u)$ 
is (the same for all actions) given by the solution of the relation
\begin{equation} \label{uprime_infty} 
  f_0(u)=f_0(u^{\prime}_\infty)+\frac{1}{2\pi}\ln2 \,,
\end{equation}
where the function $f_0(u)$ is specified in eq.~\eqref{f0_u}.
The lattice artifacts depend on the lattice actions; the leading
artifacts are specified by the function $\nu(u,z)$ where,
as will be seen later, it is convenient to express 
the $\ln L$ dependence through $z$ defined by
\begin{equation} \label{zuL}
  z = z(u,L) =f_0(u)+\frac{1}{2\pi}\ln L \,.
\end{equation}

Considering first the standard action, 
the leading lattice artifacts are linear in $z$:
\begin{equation} \label{nu_st}
  \nu_{\mathrm{std}}(u,z) = t_0(u)+t_1(u) z\,,
\end{equation}
where the functions $t_0(u), t_1(u)$ are specified in 
eqs.~\eqref{t0_st}, \eqref{t1_st}.
The coefficient $t_1(u)$ is positive, so the
asymptotic approach to the continuum limit is always from above. 
Numerical evaluation shows that $t_0(u)$ is negative for all $u$, 
hence the two terms compete with each other.
It turns out that $\nu_{\mathrm{std}}(u,L)$ is positive at $u > 0.6357$ 
for any $L$.
Below this value the approach for reasonable $L$
seems (misleadingly) to be from below. For small $u$ the asymptotic 
behavior sets in only at very large $L/a$, 
e.g. at $u=0.3$ the coefficient $\nu_{\mathrm{std}}(u,L)$ 
becomes positive only for $L/a \gtrsim 10^{5}$.

Turning to the constrained action, one can calculate 
the infinite volume correlation length $\xi$ for this case.
For large $L$ one gets from eqs.~\eqref{I1eq}, \eqref{I1b}
\begin{equation} \label{epsz}
  \frac{1}{2\epsilon} = z + \cO\left(1/L^2\right)
  \,.
\end{equation}
Using the asymptotic expression for large $u=m(L)L \approx M L$ 
\begin{equation}  \label{f0as}
  f_0(u) = -\frac{1}{2\pi} \ln u + \frac{5}{4\pi} \ln 2
  + \cO\left( \mre^{-u}\right) \,,
\end{equation}
one obtains 
\begin{equation} \label{epsxi}
  \frac{1}{2\epsilon}=\frac{1}{2\pi}\ln(\sqrt{32}\,\xi)
  \left[ 1 + \cO\left(\xi^{-2}\right)\right] \,.
\end{equation}
This is the same relation as for the standard action 
\cite{Lue82c} when the coupling $f$ is replaced by $2\epsilon$.

\begin{figure} 
  \centering
  \includegraphics[width=0.8\linewidth]{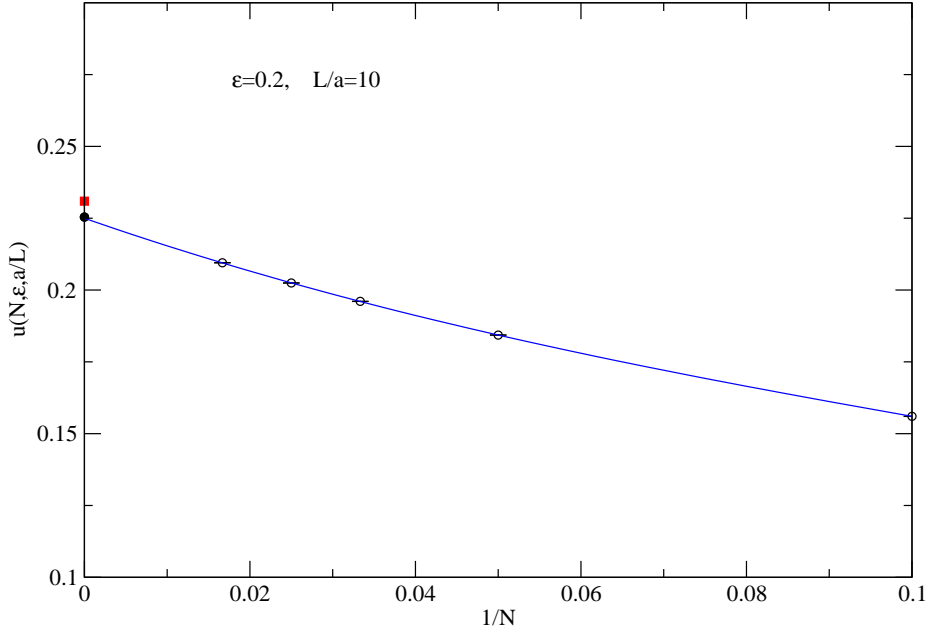}
  \caption{ $u(\epsilon,L)$ for the $\mrO(N)$ topological action
    at large $N$ values. The line is a cubic fit to the 5 data
    points. The two points at $1/N=0$ show the results obtained
    from solving the gap equations numerically. 
    The lower point is obtained from eqs.~\eqref{gap1}, \eqref{gap2}.
    For comparison we also give the result for the ``isotropic 
    action'', eq.~\eqref{gap_is}, upper point.}
  \label{MC}
\end{figure}

First, one can solve the coupled equations determining 
the step scaling function for the constrained action
numerically as indicated in appendix~\ref{AppB1a}.
We checked the numerical solution by direct Monte Carlo 
measurements for the constrained topological action ($f=\infty$) 
at $\epsilon=0.2$, $L=10$ for $N=10, 20, 30, 40, 60$.
The results are shown in figure~\ref{MC}.
The value extrapolated to $N=\infty$ is $0.225(1)$.
It agrees with the solution of the coupled gap equations,
$u(0.2,10)=0.225410$. 
For the anisotropy parameter in this case one obtains
$\rho=0.950485$. 
(Note that for the isotropic model, eq.~\eqref{gap_is}
one gets a slightly different value, $u(0.2,10)=0.230919$.)

One can also proceed analytically (see appendix~\ref{AppB1b})
and one finds that the lattice artifact $\nu_{\mathrm{con}}(u,L)$ is
now quadratic in $z$:
\begin{equation} \label{nu_C}
  \nu_{\mathrm{con}}(u,z) = \bar{t}_0(u)+\bar{t}_1(u) z +\bar{t}_2(u) z^2 \,,
\end{equation}
where the functions $\bar{t}_i(u)$ are given in eq.~\eqref{tbar_C}.
Comparing this to the result for the standard action, one has
\begin{equation}
  \bar{t}_2(u)=-8 t_1(u) \,,
\end{equation}
i.e. the asymptotic approach of $u'$ to its asymptotic value for the
constrained action is always opposite to that of the standard action, 
so that the approach in this case is always from below.
This behavior is essentially the same as observed for the $\mrO(3)$ case.

We remark that the $\ln^2L$ behavior found above is not in contradiction
to the analysis of ref.~\cite{Bal10a}, since the constrained action 
does not belong to the class of actions considered there.

\subsection{Mixed action in the large \boldmath $N$ limit }
\label{N_inf_MA}

Here we consider the mixed action
\begin{equation} \label{AS}
  S_\mathrm{mix}[\evec] =  \frac{\beta}{2} \sum_{x,\mu} (\p_\mu \evec_x)^2
  +\frac{\gamma}{4} \sum_{x,\mu} [(\p_\mu \evec_x)^2]^2 \,.
\end{equation}
It has the advantage over the constrained action that the cutoff
effects can be changed continuously, and in particular, tuned
to zero for a given quantity, similar to the $\mrO(3)$ case
discussed in section~3.

We shall take the large $N$ limit as
\begin{equation} \label{bg}
  \beta = \frac{N}{f} \,,\qquad 
  \gamma = \frac{2N}{\kappa^2} \,.
\end{equation}
The $1/N$ expansion is obtained similarly to the case 
of the constrained action and is described in appendix~\ref{AppB2}.

Introducing the effective coupling $\hat{f}=\hat{f}(f,\kappa)$ by
\begin{equation}   \label{gfk}
   \frac{1}{\hat{f}} = 
   \frac{1}{2f}+\sqrt{\frac{1}{4f^2}+\frac{1}{\kappa^2}} \,,
\end{equation}
similarly to the constrained case (cf. eqs.~\eqref{epsz}--\eqref{epsxi})
one finds for the infinite volume correlation length 
\begin{equation} \label{xig}
  \frac{1}{\hat{f}}=\frac{1}{2\pi}\ln(\sqrt{32}\,\xi)
  \left[ 1 + \cO\left(\xi^{-2}\right)\right] \,.
\end{equation}
Besides $\hat{f}$ we introduce
\begin{equation} \label{rqdef} 
  r = \frac{\kappa}{f} \,, \qquad
  \mathrm{and} \quad
  q=\frac12\left( r+\sqrt{r^2+4}\right) \,.
\end{equation}

The fixed points in the coupling space are those where $\hat{f}(f,\kappa)=0$,
i.e. the boundaries of the first quadrant of the $f$, $\kappa$
plane. The continuum limit of the step scaling function, 
$u'_\infty=\sigma(2,u)$ is of course 
universal, but the cutoff effects depend on the ratio $r$,
in general on the particular path $r(\hat{f})$
along which one approaches to the continuum limit.

In appendix~\ref{AppB2} it is shown 
that the leading artifact is of the form
\begin{equation}\label{nu_mixed}
  \nu_{\mathrm{mix}}(u,z)=  T_0(u)+ T_1(u) z + T_2(u) z^2 \,,
\end{equation}
with functions $T_i(u)$ specified in eqs.~\eqref{T_M}, \eqref{Phi_M}.

One notes that the coefficients appearing in $T_0(u)$ and $T_2(u)$ 
depend on the ratio $r$. For $r=\mathrm{constant}$ the
cutoff coefficient $\nu_\mathrm{mix}(u,z)$ is in general a 
second order polynomial in $z$ (i.e. in $\ln L$, cf. eq.~\eqref{zuL}).
The exception is the standard action ($r=\infty$) where it is a first
order polynomial. 
In this case one recovers eqs.~\eqref{nu_st} and \eqref{t1_st}, as expected.
The purely quartic action is obtained by setting $r=0$.

\begin{figure}
  \centering
  \includegraphics[width=0.8\linewidth]{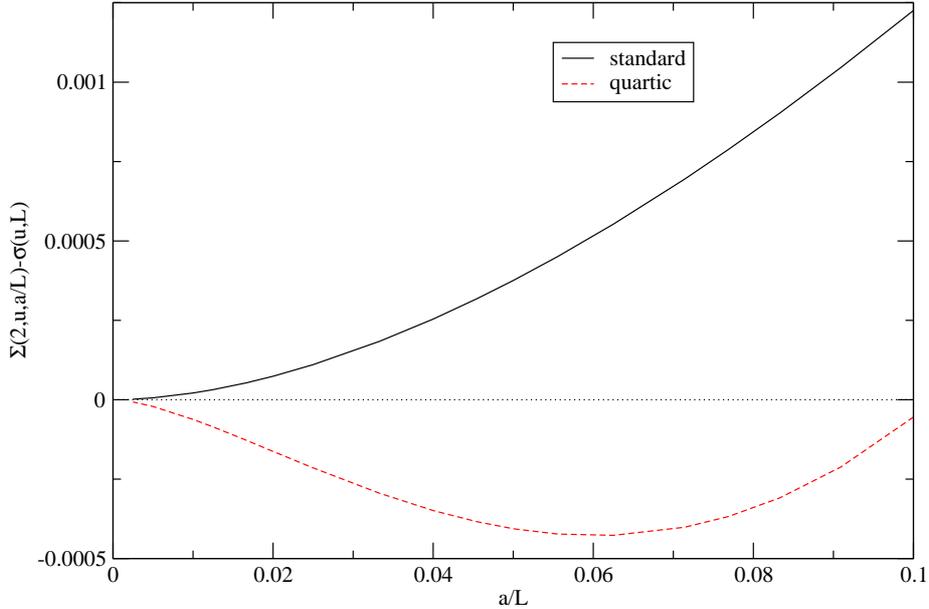}
  \caption{Deviation of $\Sigma(2,u,a/L)$ from the exact value 
    for the standard and quartic action.}
  \label{std-qrt.fig}
\end{figure}

The leading cutoff effect is positive for the standard action, 
and (for fixed $r$) negative in other cases.
Figure~\ref{std-qrt.fig} shows the deviation $\Sigma(2,u,a/L)-\sigma(2,u)$
for the standard and quartic action (with $\kappa=\infty$ and
$f=\infty$, respectively.)

As one can see from eqs.~\eqref{rqdef}, \eqref{Phi_M},
for large $z$ one can cancel
the artifact by choosing $r$ a function of $\hat{f}$ so that asymptotically 
$q^2 = 8z + \cO(1)$. This gives for large $L$ the optimal path
\begin{equation} \label{kfopt}
   \frac{\kappa^2}{2f} = \frac{\beta}{\gamma} \approx 4 \,.
\end{equation}

Note that the cancellation between the $z$ and $z^2$ terms 
takes place for all values of $u$ and scale factors $s$, hence
the lattice artifact $\Sigma(s,u,a/L)-\sigma(s,u)$
is expected to be significantly reduced for all $s$ and $u$
once it is removed, say, for $s=2$, $u=1$.

\begin{figure} 
  \centering
  \includegraphics[width=0.6\linewidth]{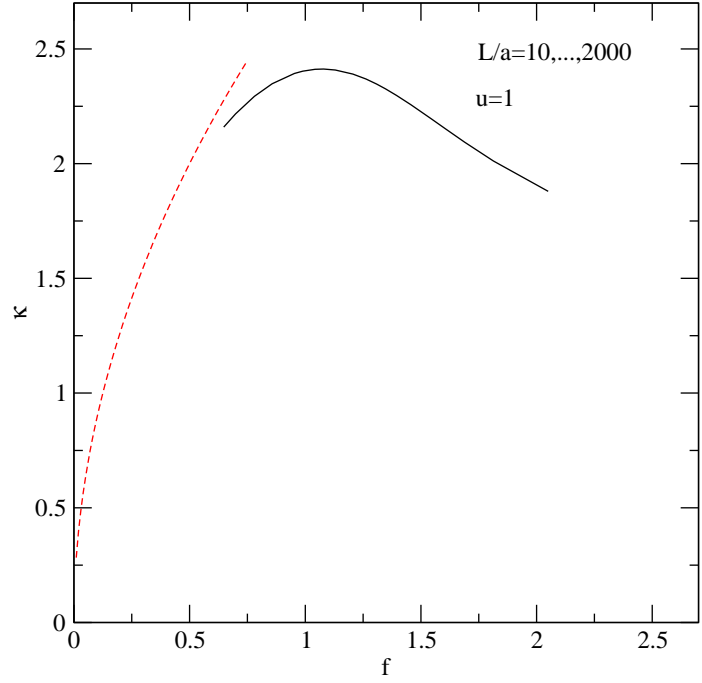}
  \caption{The curve of the optimal couplings determined 
    from $\Sigma(2,u,a/L)$ at $u=1$ for $N=\infty$. 
    The solid curve shows the results for $L/a=10, \ldots, 2000$, 
    the dashed line shows the asymptotic dependence, $\kappa = \sqrt{8 f}$.}
  \label{mixed_fkappa}
\end{figure}

\begin{figure} 
  \centering
  \includegraphics[width=0.8\linewidth]{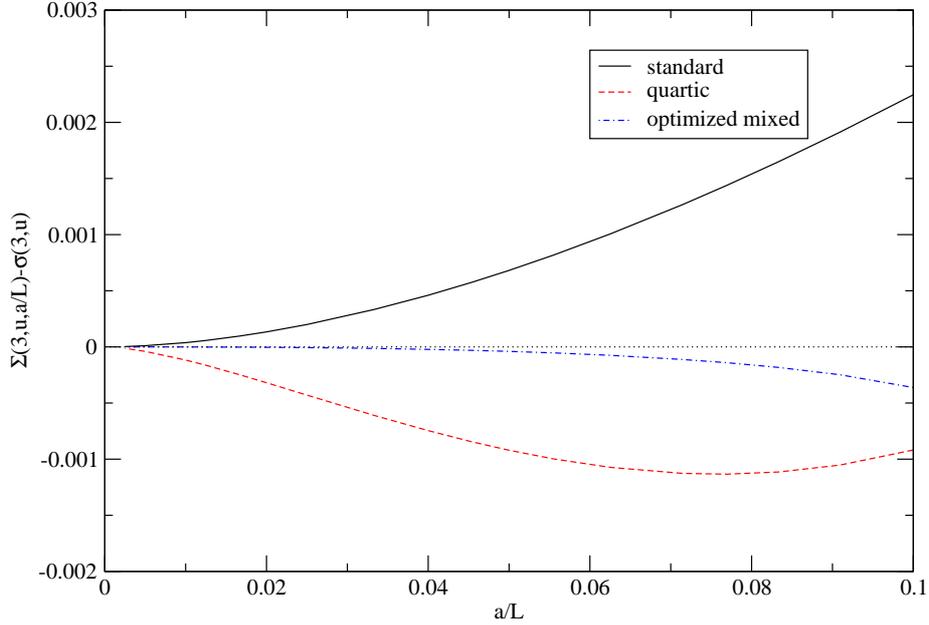}
  \caption{Deviation of $\Sigma(3,u,a/L)$
    from the exact value for the standard, quartic
    and optimized mixed action. For the latter the couplings are
    determined to have $\Sigma(2,u,a/L)=\sigma(2,u)$.}
  \label{Sigma3.fig}
\end{figure}

The optimal path in the  $f$, $\kappa$ plane shown in 
figure~\ref{mixed_fkappa} is determined from the condition 
$\Sigma(2,u,a/L)=\sigma(2,u)$ using eqs.~\eqref{I1a}--\eqref{gap3}.
The dashed line corresponds to $\kappa^2/(2f)=4$.
Figure~\ref{Sigma3.fig} shows the deviation
$\Sigma(3,u,a/L)-\sigma(3,u)$ for the optimized values $f$, $\kappa$.

\subsubsection{The renormalized 4-point coupling}
\label{N_inf_MA_gR}

Here we consider the renormalized 4-point coupling at zero
momentum, $\gR$, defined through the Binder cumulant:
\begin{align}\label{gRdef}
\gR&=-\frac{VM^2}{\Sigma^2}\sum_{x,y,x',y'}
\langle \evec_{x}\cdot \evec_{y} \; \evec_{x'}\cdot \evec_{y'}\rangle_c\,,
\\
\label{Sigmadef}
\Sigma&=\sum_{x,y}\langle \evec_{x}\cdot \evec_{y}\rangle\,,
\end{align}
and $M$ is some renormalized mass:
\begin{equation}
M=m_0+\cO(1/N)\,.
\end{equation}
Since $\gR=\cO(1/N)$ for large $N$ one needs a 
calculation at order $1/N$ which is described in appendix~\ref{AppB2a}.

Taking the infinite volume limit and the continuum limit, 
the lattice artifacts are given by 
(see appendix~\ref{AppB2b})\footnote{Here we re-introduce the 
  lattice spacing $a$.}
\begin{equation} \label{gRart}
  N\gR(aM)=8\pi+a^2 M^2  A(q,{\mathcal L}) +  \cO(a^4 M^4) \,,
\end{equation}
where ${\mathcal L}=-\ln(M^2/32)$ and
\begin{equation} \label{Aqw}
  A(q,{\mathcal L}) = 2\pi +\frac{2}{1+q^2}
  - \left(\pi+\frac{4}{1+q^2}\right){\mathcal L}
  + \frac{2}{1+q^2} {\mathcal L}^2 \,.
\end{equation}
Here $q$ is given by eq.~\eqref{rqdef}.

We see that the artifacts in $\gR$ behave very similarly to the step
scaling function artifacts (except that they are of opposite sign)
in that the leading artifact is $\cO(a^2M^2)$ times 
a second order polynomial in $\ln(aM)$.

Approaching the continuum limit along the curve $\kappa^2/(2f)=4$ 
(cf. eq.~\eqref{kfopt}), for which the logarithmic terms in the step
scaling function cancel, we have $q^2=1+8w$, and 
a similar cancellation occurs here. 

The numerical evaluation of $N \gR$ in the
infinite volume limit is described in appendix~\ref{AppB2c}.
Figure~\ref{gRN} shows
the deviation $N \gR(aM)-8\pi$ as a function of $aM$.

\begin{figure} 
  \centering
  \includegraphics[width=0.8\linewidth]{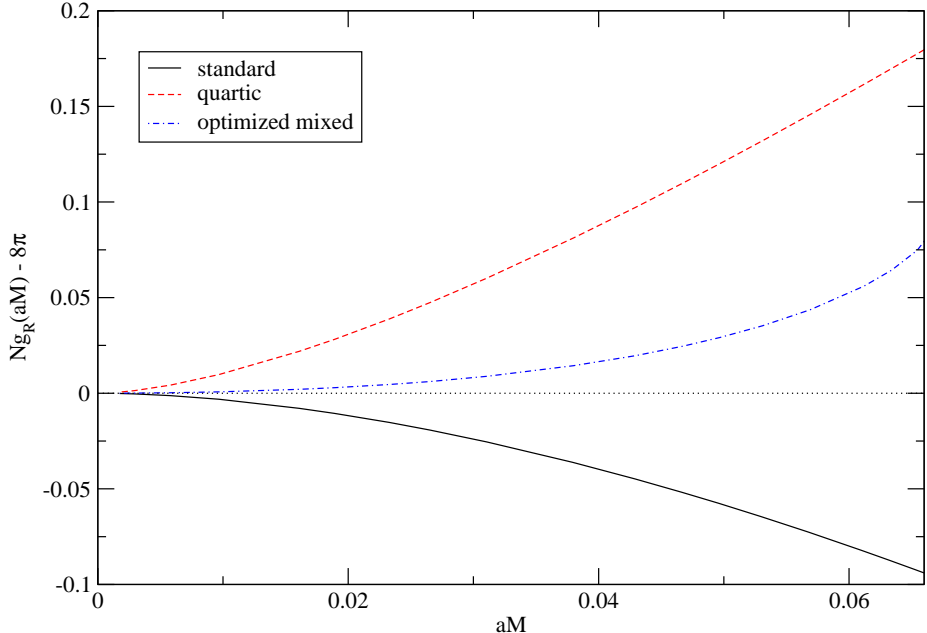}
  \caption{$N \gR-8\pi$ for the standard, quartic and optimized mixed
    action}
  \label{gRN}
\end{figure}

\section{Conclusions}
\label{Conclusions}

We have investigated a variety of lattice actions in numerical simulations of 
2-d $\mrO(3)$, $\mrO(4)$, and $\mrO(8)$ models, as well as analytically 
at $N = \infty$. 
By optimizing the constraint angle $\delta$ or the quartic coupling $\gamma$ in
very simple nearest-neighbor actions that suppress large field fluctuations, we 
have obtained a class of lattice actions with extremely small cutoff effects, 
often in the fraction of a per mille range. The simplicity of these actions 
makes them very useful for numerical simulations. In particular, the Wolff 
cluster algorithm is applicable in a straightforward manner.

It is pointed out in \cite{Bie10} and also observed here, 
that in order to belong to the corresponding universality 
class an action is not required to have the standard classical 
continuum limit.

Interestingly, both the constrained action and the mixed action
have  unexpected cutoff effects at $N = \infty$.
In particular, the leading artifact in both cases is
$\sim a^2 \log^2(aM)$ while for the class of lattice actions
considered in \cite{Bal10a} in the framework of Symanzik's effective 
theory
the leading artifact is $\sim a^2 \log(aM)$.

Although this might appear as a puzzle at first sight, in fact
Symanzik's effective theory approach is valid also
for the actions considered here. In this framework the lattice
artifacts are described by the effective continuum Lagrangian
\begin{equation}
{\cal L}_{\rm arti}=a^2\,C\,O_4
\end{equation}
where the c-number coefficient $C$ is coupling-dependent and
$O_4$ is a local O$(N)$-invariant operator. (In fact the
complete artifact Lagrangian is a linear combination of
several terms of this form.) Approaching the continuum limit
along a given curve in coupling space, $C$ becomes a function
of the inverse correlation length $aM$. On the other hand the
matrix elements of the operator $O_4$ are cutoff-independent
but depend on the physical parameters ($u$ for the case of
the step-scaling function). We know how
to calculate the asymptotic form of the dependence on $aM$ only
if coupling constant perturbation theory is applicable.

The actions considered in \cite{Bal10a} were perturbative ---
close to the continuum limit the fluctuation of the corresponding field
was determined mainly by the quadratic part of the action. 
On the other hand, for the constrained action 
(for sufficiently small $\epsilon$) these
fluctuations are restricted by the constraint, while for the mixed
action (for sufficiently large $\gamma$) by the quartic term 
coming from $\gamma (\p_\mu \evec)^4$,
hence the situation is non-perturbative. (This is obvious for the 
topological action, which is zero for the allowed configurations.)

The large $N$ results suggest that the coefficient $C$ receives
non-perturbative contributions, which behave as
$ \propto \log^2(aM)$ for $N\to\infty$.
This scenario is supported by the fact that for these
actions the $\log^2(aM)$ and the $\log(aM)$ terms have the same
$u$-dependence (namely $u^2$, cf. eqs.~\eqref{Phi_C}, \eqref{Phi_M})
as the leading term for the perturbative actions (coming from the
same operator).
This issue deserves further investigation.

At large $N$ for the standard action the squared fluctuation 
of the relative angle is roughly given by $f$, 
while in the case of the mixed action for the purely quartic case 
($\beta=0$, i.e. $f=\infty$) it is given by $\kappa$.
If by taking the continuum limit the ratio $r=\kappa/f$ approaches
infinity, then the fluctuations are dominated by the quadratic part 
of the action. In this case one observes indeed an $a^2 \log(a)$ artifact,
as for the other perturbative actions.
In particular, this is the case when one moves along the optimal curve
$\kappa=\sqrt{8f}$, where even the leading $a^2\log(a)$ artifact cancels.
By taking the continuum limit along a straight line,
$r=\mathrm{const}$ both the quadratic and the quartic terms
are relevant.\footnote{This is also reflected by the expression
for effective coupling $\hat{f}$, eq.~\eqref{gfk}.} 
This case cannot be treated in perturbation
theory, and produces a leading $a^2 \log^2(a)$ artifact.

It is remarkable that a very simple nearest-neighbor action can reduce cutoff 
effects that drastically, and it is obvious to ask whether this success extends 
to other interesting asymptotically free lattice field theories, including 2-d 
$\CP(N-1)$ models, 4-d Yang-Mills theories, or even QCD. Optimized actions are 
straightforward to construct in all these cases, and investigations in this
direction are currently in progress.
Of course, there is no exact result that helps finding the optimalized
values of the parameters in these cases. However, as was found here,
determining the optimal choice of the parameters by calculating one
physical quantity very precisely leads to an action where lattice 
artifacts
are small for a large class of other physical quantities as well.

\section*{Acknowledgments}

We like to thank C.\ Destri for useful discussions. This work is supported in 
part by funds provided by the Schweizerischer Nationalfonds (SNF). 
The ``Albert Einstein Center for Fundamental Physics'' at Bern University 
is supported by the ``Innovations- und Kooperationsprojekt C-13'' 
of the Schweizerische Uni\-ver\-si\-t\"ats\-kon\-fe\-renz (SUK/CRUS).
J.~B. and F.~N. thank the MPI Munich, where part of this work has been
done, for hospitality. 
This investigation has been supported in part by the Hungarian
National Science Fund OTKA (under K 77400) 
and by the Regione Lombardia and CILEA Consortium
through a LISA 2011 grant.

\begin{appendix}

\section{Analytic Study of Cutoff Effects in the 1-d 
  \boldmath $\mrO(3)$ Model}
\label{AppA}

For the constrained action in one dimension 
the eigenvalues of the transfer matrix 
(with a properly chosen normalization factor) are given by
\begin{equation} \label{lambda_n}
  \lambda_n(\beta,c) = \int_c^1 \mrd z\, \mre^{\beta(z-1)} P_n(z)
\end{equation}
The standard action corresponds to $c=\cos\delta=-1$.

With the notation $x=1/\beta$, $\epsilon=\exp((c-1)\beta)$:
\begin{equation} \label{lambda_0}
  \lambda_0 = x - \epsilon x \,,
\end{equation}

\begin{equation} \label{lambda_1}
  \lambda_1 = x-x^2 - \epsilon (c x- x^2) \,,
\end{equation}

\begin{equation} \label{lambda_2}
  \lambda_2 = x - 3 x^2 + 3 x^3  - \epsilon
  \left(
    \frac12 (3 c^2 - 1 ) x - 3 c x^2 + 3 x^3 
  \right) \,,
\end{equation}

\begin{multline} \label{lambda_3}
  \lambda_3 = x - 6 x^2 + 15 x^3- 15 x^4  \\
  - \epsilon
  \left(
    \frac12 (5 c^3 -3 c) x - \frac12 (15 c^2 -3) x^2 
    + 15 c x^3 - 15 x^4
  \right)  \,.
\end{multline}

The excitation energies of the rotator (in lattice units) are
\begin{equation} \label{E_n}
  a E_n = -\log\left( \frac{\lambda_n}{\lambda_0}\right)\,.
\end{equation}
The lattice spacing is determined here by $a E_1= 1$.

The ratios of the energies are given by
\begin{equation} \label{r_n}
  r_n = \frac{E_n}{E_1}\,.
\end{equation}

In the continuum limit ($c$ fixed, $\beta\to\infty$)
they approach the well-known result for the quantum
rotator,
\begin{equation} \label{r_nc}
  \lim_{\beta\to\infty}r_n = \frac12 n (n+1) \,.
\end{equation}
In particular
\begin{equation} \label{sr2}
  r_2 = 3 - x^2 -\frac52 x^3 + \ldots
\end{equation}
\begin{equation} \label{sr3}
  r_3 = 6 - 5 x^2 -\frac{25}{2} x^3 + \ldots
\end{equation}

\begin{figure}[t]
  \centering
  \vskip1cm
  \includegraphics[width=0.8\textwidth,angle=0]{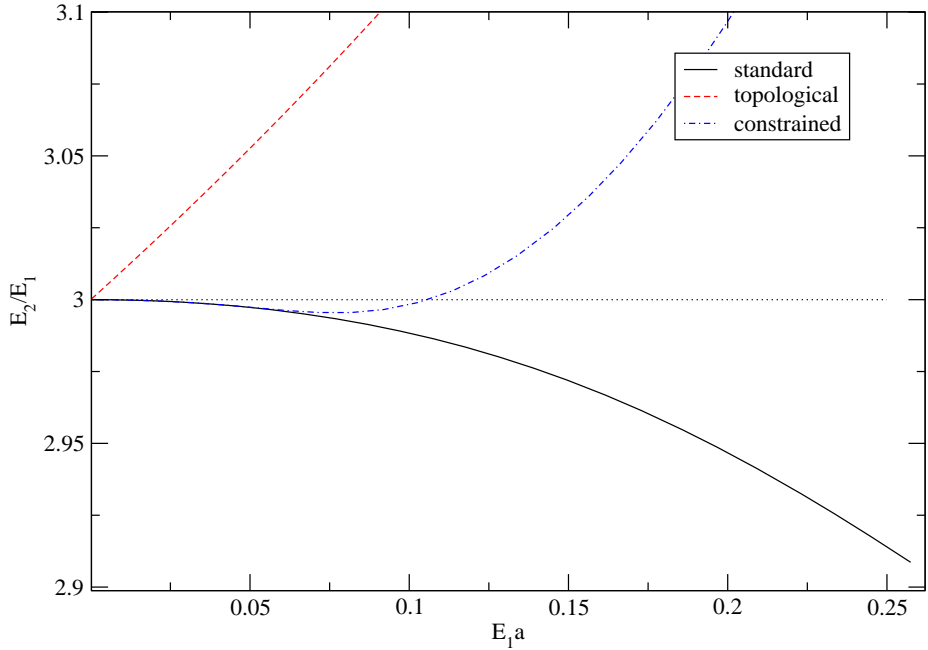}
  \caption{\it The ratio of the mass gaps for the $I = 2$ and $I = 1$ 
    states as a function of the lattice spacing for the standard, 
    topological, and constrained action. 
    For the latter the constraint is fixed at $E_1 a= 0.1$}
  \label{1d_E2}
\end{figure}

\begin{figure}[t]
  \centering 
  \vskip1cm
  \includegraphics[width=0.8\textwidth,angle=0]{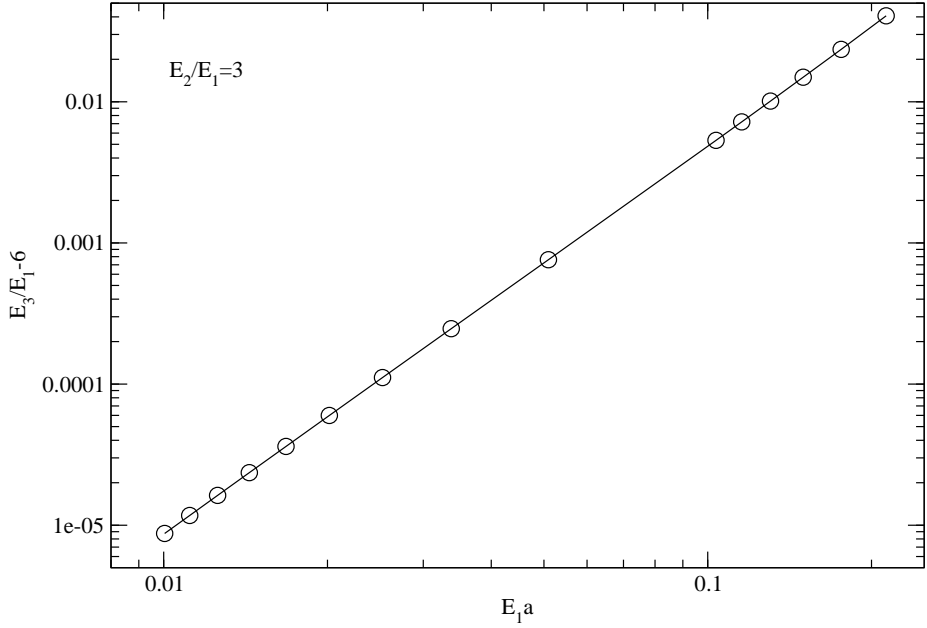}
  \caption{\it The deviation of the ratio of the mass gaps for 
    the I=3 and I=1 states from its continuum value.
    The constraint $c=c(\beta)$ is chosen so that $E_2/E_1=3$.}
  \label{1d_E3}
\end{figure}

In the case of the topological action,
$\beta=0$, $c\to 1$, one has
\begin{equation} \label{cr2}
  r_2 = 3 + \frac12 (1-c) + \frac38 (1-c)^2 + \ldots
\end{equation}
\begin{equation} \label{cr3}
  r_3 = 6 + \frac52 (1-c) + \frac{25}{8} (1-c)^2 + \ldots
\end{equation}
Comparing eqs.~\eqref{sr2},\eqref{sr3},\eqref{cr2},
\eqref{cr3} one sees that the discretization errors for
the standard action and for the the topological action
have opposite signs.
Fixing the constraint e.g. to $c=0.4$ one obtains
the resulting artifacts for $r_2$ which lie in between
those for the standard and the topological action 
(cf.~figure~\ref{1d_E2})
An interesting possibility is to choose $c=c(\beta)$ in such a way
that the first ratio is exact,  $r_2(\beta,c(\beta))=3$.
Of course, this procedure does not eliminate the discretization 
errors in other quantities,
but it turns out that it improves the convergence 
to the continuum limit.
Figure~\ref{1d_E3} 
shows $r_3(\beta,c(\beta))-6$ as a function of $a E_1$,
which in a log-log plot looks nearly linear. 
From the numerical values one finds that 
$ r_3(\beta,c(\beta))-6 \propto a^\alpha$,
with the power $\alpha\approx 2.77$.

\section{Technical Details of the \boldmath $N = \infty$ Calculation}
\label{AppB}

\subsection{Analytic behavior of the step scaling function: constrained 
action}
\label{AppB1}

Introducing
\begin{align}  \label{I1def}
  I_1(\rho,u,L) & = \frac{1}{L}\sum_p \frac{1}{\sqrt{\omega(\omega+4)}}
  \,, \\  \label{I2def}
  I_2(\rho,u,L) & = \frac{1}{L}\sum_p \frac{\omega}{\sqrt{\omega(\omega+4)}}
  \,,
\end{align}
(with $m_0$ given by eq.~\eqref{m0u}) one obtains
\begin{align}  \label{I1eq}
  I_1(\rho,u,L) & = \frac{1}{m_0^2 + (\rho + 1)\epsilon}
  \,, \\  \label{I2eq}
  I_2(\rho,u,L) & = \frac{m_0^2 + \rho\epsilon}{m_0^2 + (\rho + 1)\epsilon}
  \,.
\end{align}

The asymptotic expansion of $I_1$, $I_2$ can be obtained from the
results of Caracciolo and Pelissetto \cite{CaPe}
(expressing the expansion in terms of $u=m L$ instead of $m_0 L$ used therein).
One has asymptotic expansions
\begin{align}
  I_1(\rho,u,L) & = z +\frac{1}{L^2}B(u,z)+\dots \,, \label{I1b}\\
  I_2(\rho,u,L) & = \frac{1}{2}+\frac{1}{L^2}D(u,z)+\dots \,, \label{I2b}
\end{align}
where we recall $z$ is defined by
\begin{equation} 
  z = z(u,L) =f_0(u)+\frac{1}{2\pi}\ln L \,,
\end{equation}
and
\begin{align}
  B(u,z)& = f_1(u)+\frac18 u^2 f_0(u) -\frac18 u^2 z -\frac{\rho_1}{2}\left[
    z+uf_0^\prime(u)+\frac{1}{4\pi}\right] \,,\\
  D(u,z)& = f_2(u)-\frac12 u^2 f_0(u) + \frac12 u^2 z +\frac{\rho_1}{2\pi} \,.
\end{align}
Here $\rho_1$ appears in the expansion of $\rho(u,L)$
\begin{equation} \label{rho}
  \rho(u,L)=1+\frac{1}{L^2}\rho_1(u,z)+\cO\left( 1/L^4\right) \,.
\end{equation}
Note that the isotropy is recovered ($\rho\to 1$) for $L\to\infty$,
as one can see from eq.~\eqref{I2eq} and from the infinite volume limit
\begin{equation} \label{I2inf}
  I_2(\rho,u,\infty) = \frac{2}{\pi}\arctan\left(\sqrt{\rho}\right) \,.
\end{equation}

The remaining $L$ dependence in $\rho_1$ is assumed to be only logarithmic.

The functions $f_i(u)$ are given by
\begin{align}\label{f0_u}
  f_0(u) & = \frac{1}{2u}+\frac{1}{2\pi}
  \left[k+G_0\left(\frac{u}{2\pi}\right)\right] \,,
  \\
  f_1(u) & = \frac{\pi}{6}\left[\frac{1}{12}
    -G_1\left(\frac{u}{2\pi}\right)\right] -\frac{u}{12} - \frac{u^2}{16\pi}
  \left[k+\frac23 G_0\left(\frac{u}{2\pi}\right)-\frac13\right] \,,
  \\
  f_2(u) & = 2\pi G_1\left(\frac{u}{2\pi}\right)-\frac{\pi}{6}+\frac{u}{2}
  +\frac{u^2}{8\pi}(2k-1) \,,
  \\
  k & = \gamma_E-\ln\pi+\frac12\ln2 \,.
\end{align}

We also need the derivative of $f_0(u)$:
\begin{equation} \label{f0prime} 
  f^{\prime}_0(u)=-\frac{1}{2u^2}
  -\frac{u}{(2\pi)^3}H_1\left(\frac{u}{2\pi}\right)\,.
\end{equation}
Here\footnote{Note there is an error in eq.~(A.5) of \cite{CaPe}: 
it should read 
$G_1(\alpha)=2\sum_{k=1}^\infty\,\frac{(-1)^k}{k+1}\binom{2k}{k}
  \zeta(2k+1)\left(\frac{\alpha}{2}\right)^{2k+2}\,.$}
\begin{align}
  G_0(x) & =
  \sum_{n=1}^\infty\left[\frac{1}{\sqrt{n^2+x^2}}-\frac{1}{n}\right]\,,
  \\
  G_1(x) & = \sum_{n=1}^\infty\left[\sqrt{n^2+x^2}-n-\frac{x^2}{2n}\right]\,,
  \\
  H_1(x) & = \sum_{n=1}^\infty(n^2+x^2)^{-3/2}\,.
\end{align}
Note for further reference that $f'_0(u)<0$.

For the step scaling function we are interested in $u'(u,L)=m(2L)2L$ 
for fixed $u(L)$.

Consider first the standard action.
From the gap equation $I_1(u,L)=1/f$ and $I_1(u',2L)=1/f$
for $u'=u'(u,L)$ one has the equation
\begin{multline} \label{zLzpL_st}
   z + \frac{1}{L^2}\left[f_1(u)+\frac18 u^2 f_0(u) 
    - \frac18 u^2 z \right]
 \\
   = z' + \frac{1}{4L^2}\left[f_1(u')+\frac18 u^{\prime 2} f_0(u') 
    - \frac18 u^{\prime 2} z' \right]
  +\cO(1/L^4)\,.
\end{multline}
Expanding $u'$ as in eq.~\eqref{uprime} one sees
that the continuum value $u'_\infty$ is given
by the solution to eq.~\eqref{uprime_infty}.

From here and eq.~\eqref{zuL} one has for $z'=z(u',2L)$
\begin{equation} \label{zpz} 
  z' = z + \frac{1}{L^2} f_0'(u'_\infty) \nu(u,z)
  + \cO\left(\frac{1}{L^4}\right) \,.
\end{equation}
Because of this relation 
one can replace $z'$ by $z$ in the $1/L^2$ terms in eq.~\eqref{zLzpL_st}.
This is the reason for using $z$ instead of $\ln L$.

For the leading lattice artifact one obtains eq.~\eqref{nu_st} where
\begin{align}  \label{t0_st}
  t_0(u) & = \frac{1}{f^\prime_0(u^{\prime}_\infty)} \left[
    f_1(u)+\frac18 u^2 f_0(u)
    -\frac14 \left(f_1(u^{\prime}_\infty) + 
      \frac18 u^{\prime 2}_\infty f_0(u^{\prime}_\infty)
    \right)
  \right]\,,
  \\   \label{t1_st}
  t_1(u) & = -\frac{1}{8 f^\prime_0(u^{\prime}_\infty)} \left[u^2-\frac14
    u_\infty^{\prime2}\right]\,.
\end{align}

Note that according to eq.~\eqref{f0prime} $f'_0(u)$ is negative for all
$u$. Since $u'_\infty < 2 u$ the coefficient $t_1(u)$ is positive.

\subsubsection{Solving the coupled equations numerically}
\label{AppB1a}

Eliminating $\epsilon$ from eqs.~\eqref{I1eq},\eqref{I2eq}
one gets (with $m_0$ given by eq.~\eqref{m0u})
\begin{equation} \label{I1I2rho}
  (1+\rho) I_2(\rho,u,L) = m_0^2 I_1(\rho,u,L) + \rho\,. 
\end{equation}
This determines $\rho=\rho(u,L)$. Then one finds the corresponding
$\epsilon$ by
\begin{equation} \label{epsI1}
  \epsilon = \left( \frac{1}{I_1(\rho,u,L)}-m_0^2\right) \frac{1}{\rho+1}\,.
\end{equation}
Keeping $\epsilon$ fixed one can calculate $u'=u(\epsilon,2L)$.

\subsubsection{Analytic form of the leading artifact}
\label{AppB1b}

Inserting the asymptotic expansions \eqref{I1b},\eqref{I2b},\eqref{rho}
we get
\begin{equation}
  \rho_1(u)=\frac{2\pi}{\pi-2}\left( 2 f_2(u)-u^2 f_0(u)\right)\,,
\end{equation}
which is independent of $z$ and
\begin{equation} \label{epsinv_C}
  \frac{1}{2\epsilon} =
   \frac{\rho I_1}{2(I_2-m_0^2 I_1)}\\
   = z + \frac{1}{L^2}\left( \Phi_0(u) + \Phi_1(u) z + \Phi_2(u) z^2  \right)
   +\ldots \,,
\end{equation}
where
\begin{equation} \label{Phi_C}
  \begin{split}
    \Phi_0(u) & = f_1(u) + \frac18 u^2 f_0(u) 
    -\frac12 \rho_1(u)\left( u f'_0(u) + \frac{1}{4\pi}\right) \,, \\
    \Phi_1(u) & =  - \frac18 u^2 \,, \\
    \Phi_2(u) & = u^2 \,.
  \end{split} 
\end{equation}

One now obtains eq.~\eqref{nu_C} with
\begin{equation} \label{tbar_C}
  \bar{t}_i(u)  = \frac{1}{f'_0(u'_\infty)}
  \left(\Phi_i(u) - \frac14 \Phi_i(u^{\prime}_\infty)\right)\,.
\end{equation}

\subsection{\boldmath $1/N$ expansion for the mixed action}
\label{AppB2}

Here we discuss the mixed action given in eq.~\eqref{AS}
in the large $N$ limit given by eq.~\eqref{bg}.

The partition function is
\begin{equation} \label{Z1b} 
  Z = \int_{\evec} \exp\Big\{ - S_\mathrm{mix}[\evec] \Big\}
  \prod_x \delta( \evec_x^2-1) \,.
\end{equation}

Introducing the auxiliary variables $\alpha_x$, $\eta_{x\mu}$ one obtains
\begin{equation} \label{Aeff1}
  S_{\mathrm{eff}}[\evec] =  \frac12 \sum_{x,\mu} 
\left(\beta + 2i \sqrt{\gamma} \eta_{x\mu} \right) (\p_\mu \evec_x)^2
+\sum_{x,\mu} \eta_{x\mu}^2 - i \sum_x \alpha_x (\evec_x^2-1) \,.
\end{equation}

Rescaling and shifting the integration contour as
\begin{equation}
  \alpha_x \to \frac12 N \left( 
    i h + \frac{\alpha_x}{\sqrt{N}}
  \right)
 \,,\qquad 
  \eta_{x\mu} \to \sqrt{\frac{N}{2}}
  \frac{\kappa}{2} \left(
    -i v_\mu + \frac{\eta_{x\mu}}{\sqrt{N}}
\right) \,,
\end{equation}
one gets
\begin{multline}  \label{Aeff3}
  S_{\mathrm{eff}}[\evec] =  \frac12 N 
  \left[ \sum_{x,\mu} \left( \frac{1}{f} + v_\mu
      +i\frac{\eta_{x\mu}}{\sqrt{N}}\right) (\p_\mu \evec_x)^2 \right.
  \\
  \left.
    + \frac{\kappa^2}{4} \sum_{x,\mu}
    \left( -i v_\mu + \frac{\eta_{x\mu}}{\sqrt{N}} \right)^2
    + \sum_x \left(h- i \frac{\alpha_x}{\sqrt{N}}\right) (\evec_x^2-1)
    \right] \,.
\end{multline}
Note that one has to allow for an anisotropy in $v_\mu$ because
the lattice is not cubic, $L_t \ne L_s$.

After integrating out the spin fields we get an effective action in the
auxiliary fields
\begin{multline} \label{Aeffbar_m}
  \overline{S}_\mathrm{eff} = \frac{N}{2}\left[ \mathrm{tr}\ln R -
    V h -  V \frac{\kappa^2}{4} \sum_\mu v_\mu^2
  \right] \\
  - i\frac{\sqrt{N}}{2} \Big[ -\sum_x \alpha_x 
  + \frac{\kappa^2}{2} \sum_{x,\mu} v_\mu \eta_{x\mu} \Big] 
  + \frac{\kappa^2}{8} \sum_{x,\mu}\eta_{x\mu}^2 \,,
\end{multline}
with
\begin{align}
  & R_{xy}=S_{xy}+\frac{i}{\sqrt{N}}T_{xy} \,,
  \\
  & S_{xy} = h \delta_{xy} +\sum_\mu w_\mu
  \left[2\delta_{xy}-\delta_{y,x+\hat{\mu}}-\delta_{y,x-\hat{\mu}}\right]\,,
  \\
  & T_{xy} = -\alpha_x\delta_{xy} + t_{xy}\,,
  \\
  & t_{xy} = \sum_\mu
  \left[\eta_{x\mu}\left(\delta_{xy}-\delta_{y,x+\hat{\mu}}\right)
    +\eta_{(x-\hat{\mu})\mu}\left(\delta_{xy}-\delta_{y,x-\hat{\mu}}\right)
  \right] \,,\\
  & w_\mu = \frac{1}{f}+v_\mu \,.
\end{align}

For the inverse $S^{-1}_{xy}$, and $D(p)$ we have the previous expressions,
eqs.~\eqref{Sinv},\eqref{Dp}.

In the leading order of $1/N$ expansion
one obtains the following equations for $h$ and $v_\mu$
\begin{align} \label{gap01m}
  & \frac{1}{V} \sum_p \frac{1}{D(p)} = 1 \,,
  \\
  \label{gap02m}
  & \frac{1}{V} \sum_p \frac{K_\mu(p)}{D(p)} = \frac12 \kappa^2 v_\mu \,.
\end{align}
This yields the relation
\begin{equation} \label{gap03m}
  \frac12 \kappa^2 \sum_\mu v_\mu w_\mu + h = 1 \,.
\end{equation}

For the $d=2$ case, with the same notations 
as in the constrained case (cf.~appendix~\ref{AppB1})
we get for $L_t=\infty$: 
\begin{align}  \label{I1a}
  I_1(\rho,u,L) & = w_0  \,, \\ 
  \label{I2a}
  I_2(\rho,u,L) & = w_0 \left( \frac{\kappa^2}{2} \rho v_1 + m_0^2\right)
  \,,
\end{align}
and
\begin{equation} \label{gap3}
  \frac12 \kappa^2 (v_0 + \rho v_1) + m_0^2 = \frac{1}{w_0} \,.
\end{equation}

Eliminating $v_0$, $v_1$ from eqs.~\eqref{I1a},\eqref{I2a} and \eqref{gap3}
we can express the two couplings in terms of $u$,
$L$ and the remaining saddle point parameter, $\rho$:

\begin{align}
  \frac{2}{\kappa^2} & = \frac{\rho(\rho-1)}{(1+\rho)I_2-\rho-m_0^2I_1}I_1^2 \,,
  \label{keq}\\
  \frac{1}{f}& = I_1 + \frac{2}{\kappa^2}\cdot \frac{I_2-1}{I_1}
  \,.
  \label{feq}
\end{align}

Inserting the asymptotic expansions \eqref{I1b}, \eqref{I2b} 
into eq.~\eqref{feq} one obtains
\begin{equation}  \label{lead}
  \frac{1}{f} = z - \frac{1}{\kappa^2z} + \cO\left( \frac{1}{L^2}\right) \,.
\end{equation}
Solving this for $z$ one finds
\begin{equation}
  z(u,L) = \frac{1}{\hat{f}} + \cO\left( \frac{1}{L^2}\right) \,,
  \label{zg}
\end{equation}
where the effective coupling $\hat{f}(f,\kappa)$ is defined in eq.~\eqref{gfk}.

Inserting the asymptotic expansions\footnote{Using 
eqs.~\eqref{keq} and \eqref{I2inf} one can show that 
$\rho(u,\infty)=1$.}
\eqref{I1b}, \eqref{I2b}, and \eqref{rho} into eq.~\eqref{keq},
and using eq.~\eqref{zg} one obtains
\begin{equation} \label{rho1}
  \rho_1(u) = 
  \frac{2(2f_2(u)-u^2f_0(u))}{1-\frac{2}{\pi}+q^2} \,,
\end{equation}
up to $\cO(L^{-2})$. 
Here $q$ is a function of the coupling ratio $r=\kappa/f$
given by eq.~\eqref{rqdef}.
Note that taking $r=\infty$ one should recover the result for 
the standard action, while taking $r=0$ that one for 
the pure quartic action.

We now eliminate $I_2$ using eq.~\eqref{keq} and obtain from 
eq.~\eqref{feq}
\begin{equation}
  \frac{1}{f}=\frac{1}{1+\rho}\left\{(1+\rho^2)I_1
    -\frac{2}{\kappa^2I_1}+\frac{2m_0^2}{\kappa^2}\right\} \,.
\end{equation}
Inserting the asymptotic expansions one obtains
\begin{multline}
  \frac{1}{f}= z - \frac{1}{\kappa^2 z}  
  +\frac{1}{L^2} \left\{ 
    \left[f_1(u)+ \frac18 u^2 f_0(u) - \frac18 u^2 z \right.\right.\\
  \left. \left.
      -\frac12 \rho_1(u)\left(
        uf_0^\prime(u)+\frac{1}{4\pi}\right)\right]\left(
      1+\frac{1}{\kappa^2z^2}\right)+\frac{u^2}{\kappa^2}\right\}
  + \ldots 
\end{multline}

Using these relations one obtains the leading lattice artifacts
given in eq.~\eqref{nu_mixed} with the functions $T_j(u)$ given by
\begin{equation} \label{T_M}
  T_i(u)  = \frac{1}{f'_0(u'_\infty)}
  \left(\Phi_i(u) - \frac14 \Phi_i(u^{\prime}_\infty)\right) \,,
\end{equation}
\begin{equation} \label{Phi_M}
  \begin{split}
    \Phi_0(u) & = f_1(u) + \frac18 u^2 f_0(u) 
    -\frac12 \rho_1(u)\left(u f_0^\prime(u)+
      \frac{1}{4\pi}\right) \,, \\
    \Phi_1(u) & = -\frac18 u^2 \,, \\
    \Phi_2(u) & = \frac{1}{1+q^2} u^2 \,. \\
  \end{split}
\end{equation}

\subsubsection{The renormalized 4-point coupling in leading order}
\label{AppB2a}

In order to compute at higher orders in the $1/N$ expansion
let us introduce a source in the action; 
\begin{equation}
  S_{\mathrm{eff}}[\evec,\Jvec] = S_{\mathrm{eff}}[\evec] 
    +\sum_x \Jvec_x\cdot \evec_x
\end{equation}
Then after integrating out the spin fields we get
\begin{equation}
\overline{S}_\mathrm{eff}[\Jvec]=\overline{S}_\mathrm{eff} 
-\frac{1}{2N}\sum_{x,y}\Jvec_x (R^{-1})_{xy} \Jvec_y\,.
\end{equation}

The propagators of the auxiliary fields are given by the quadratic terms
\begin{equation}
  \overline{S}_{\mathrm{eff,quadratic}} =
  \frac14 \mathrm{tr} \left(S^{-1}TS^{-1}T\right)
  +\frac{\kappa^2}{8}\sum_{x,\mu}\eta^2_{x\mu} \,,
\end{equation}
where $S^{-1}_{xy}$ is given by eq.~\eqref{Sinv}.

Defining the Fourier transforms
\begin{align}
  &\alpha_x=\frac{1}{V}\sum_p\,\mre^{ipx}\widetilde{\alpha}(p)\,,
  \\  
  &\eta_{x\mu}=\frac{1}{V}\sum_p\,
  \mre^{ip(x+\hat{\mu}/2)}\widetilde{\eta}_\mu(p)\,,
\end{align}
we have\footnote{Here $\hat{p}_\mu= 2 \sin(p_\mu/2)$.}
\begin{equation}
(TS^{-1})_{xy} = \frac{1}{V^2}\sum_{p,q}\frac{\mre^{ip(x-y)+iqx}}{D(p)}
\left\{-\widetilde{\alpha}(q)+\widehat{(p+q)}_\mu\hat{p}_\mu
\widetilde{\eta}_\mu(q)\right\}\,,
\end{equation}
so the quadratic term in the auxiliary fields is
\begin{align}
  & \overline{S}_{\mathrm{eff,quad}} =
  \frac{\kappa^2}{8V}\sum_{q,\mu}
  \widetilde{\eta}_\mu(q)\widetilde{\eta}_\mu(-q)
  \nonumber\\
  &+\frac14 \frac{1}{V^2}\sum_{q,p} 
  \frac{\left[\widetilde{\alpha}(q)-
      \hat{p}_\mu\hat{r}_\mu\widetilde{\eta}_\mu(q)\right]
    \left[\widetilde{\alpha}(-q)-
      \hat{p}_\nu\hat{r}_\nu\widetilde{\eta}_\nu(-q)\right]}
  {D(p)D(r)}\,,\,\,\,\,r=p+q\,.
\end{align}

Define ($r=p+q$):
\begin{align}
  H(q)&=\frac{1}{V}\sum_{p}\frac{1}{D(p)D(r)}\,,
  \\
  H_\mu(q)&=\frac{1}{V}\sum_{p}\frac{\hat{p}_\mu\hat{r}_\mu}{D(p)D(r)}\,,
  \\
  H_{\mu\nu}(q)&=\frac{1}{V}\sum_{p}
  \frac{\hat{p}_\mu\hat{r}_\mu\hat{p}_\nu\hat{r}_\nu}{D(p)D(r)}\,.
\end{align}
Note all $H-$functions are even in $q$.
Then the leading order quadratic term can be written
\begin{align}
  \overline{S}_{\mathrm{eff,quad}} &=
  \frac14 \frac{1}{V}\sum_q  
  \Bigl[ \widetilde{\alpha}(q)H(q)\widetilde{\alpha}(-q)
  -\widetilde{\alpha}(q)H_\mu(q)\widetilde{\eta}_\mu(-q)
  -\widetilde{\alpha}(-q)H_\mu(q)\widetilde{\eta}_\mu(q)
  \nonumber\\
  &+\widetilde{\eta}_\mu(q)H_{\mu\nu}(q)\widetilde{\eta}_\nu(-q)\Bigr]
  +\frac{\kappa^2}{8V}\sum_{q,\mu}\widetilde{\eta}_\mu(q)
  \widetilde{\eta}_\mu(-q)\,.
\end{align}
Defining
\begin{equation}
  \widetilde{\beta}(q)=\widetilde{\alpha}(q)
  -\frac{H_\mu(q)}{H(q)}\widetilde{\eta}_\mu(q)\,,
\end{equation}
we have diagonalized the quadratic part:
\begin{align}
  \overline{S}_{\mathrm{eff,quad}} &=
  \frac14 \frac{1}{V}\sum_q  
  \Bigl[\widetilde{\beta}(q)H(q)\widetilde{\beta}(-q)   
  \nonumber\\
  &+\widetilde{\eta}_\mu(q)
  \left\{H_{\mu\nu}(q)-\frac{H_\mu(q)H_\nu(q)}{H(q)}
    +\frac12 \kappa^2 \delta_{\mu\nu}\right\}
  \widetilde{\eta}_\nu(-q)\Bigr]\,,
\end{align}

\begin{equation}
  (S^{-1}TS^{-1})_{xy} = 
  \frac{1}{V^2}\sum_{p,q}\frac{\mre^{irx-ipy}}{D(p)D(r)}
  \left\{-\widetilde{\beta}(q)
    +X_\mu(p,q)\widetilde{\eta}_\mu(q)\right\}\,,\,\,\,\,r=p+q\,,
\end{equation}
where
\begin{equation}
  X_\mu(p,q)\equiv 
  \hat{r}_\mu\hat{p}_\mu-\frac{H_\mu(q)}{H(q)}\,,\,\,\,r=p+q\,.
\end{equation}

For the connected 4-point coupling in leading order we obtain
\begin{align}
  &\langle \evec_{x}\cdot \evec_{y} \evec_{x'}\cdot \evec_{y'}\rangle_c
  =-\frac{2}{N}\frac{1}{V^3}\sum_{p,q,p'}
  \frac{\mre^{ipx-iry}}{D(p)D(r)}\frac{\mre^{ip'x'-ir'y'}}{D(p')D(r')}
  \nonumber\\
  &\times\left\{
    \widetilde{\triangle}(q)+X_\mu(p,q)X_\nu(p',q)
    \widetilde{\triangle}_{\mu\nu}(q)
  \right\}\,,\,\,\,\,r=p+q,r'=p'-q\,,
\end{align}
where
\begin{align}
  &\widetilde{\triangle}(q)=\frac{1}{H(q)}\,,
  \\
  &\widetilde{\triangle}_{\mu\rho}(q)
  \left\{H_{\rho\nu}(q)-\frac{H_\rho(q)H_\nu(q)}{H(q)}
    +\frac12 \kappa^2 \delta_{\rho\nu}\right\}
  =\delta_{\mu\nu}\,.
\end{align}

We consider the isotropic case $T=L$, $w_0=w_1=w$, then in leading order
for the 2- and 4-point functions:
\begin{align}
  &\Sigma=\frac{V}{wm_0^2}+\cO(1/N)\,,
  \\
  &\sum_{x,y,x',y'}\langle \evec_{x}\cdot \evec_{y} \evec_{x'}\cdot 
  \evec_{y'}\rangle_c
  =-\frac{2}{N}\frac{V}{(wm_0^2)^4}\left\{
    \widetilde{\triangle}(0)+\frac{H_\mu(0)H_\nu(0)}{H(0)^2}
    \widetilde{\triangle}_{\mu\nu}(0)\right\}
  \nonumber\\
  &+\cO(1/N^2)\,,
\end{align}
so that the renormalized coupling, eq.~\eqref{gRdef} is given by
\begin{equation}
  N\gR=\frac{2}{w^2M^2}\left\{
    \widetilde{\triangle}(0)+\frac{H_\mu(0)H_\nu(0)}{H(0)^2}
    \widetilde{\triangle}_{\mu\nu}(0)\right\}+\cO(1/N)\,.
\end{equation}

Note (setting to this order $m_0=M$)
\begin{align}
  H(0)&=\frac{1}{w^2}\overline{H}_0\,,
  \\
  H_\mu(0)&=\frac{1}{dw^2}\overline{H}_1\,,
  \\
  H_{\mu\nu}(0)&=\frac{1}{w^2V}\sum_p
  \frac{\hat{p}^2_\mu\hat{p}^2_\nu}{(\hat{p}^2+M^2)^2}
  \\
  &=\frac{1}{w^2d}\left\{\overline{H}_2\delta_{\mu\nu}
    +\overline{H}_3\left(1-d\delta_{\mu\nu}\right)\right\}\,,
\end{align}
where
\begin{align}
  \overline{H}_0&=\frac{1}{V}\sum_p\frac{1}{(\hat{p}^2+M^2)^2}\,,
  \\
  \overline{H}_1&\equiv\frac{1}{V}
  \sum_p\frac{\hat{p}^2}{(\hat{p}^2+M^2)^2}\,,
  \\
  \overline{H}_2 &\equiv\frac{1}{V}
  \sum_p\frac{(\hat{p}^2)^2}{(\hat{p}^2+M^2)^2}\,,
  \\
  \overline{H}_3 &\equiv\frac{1}{(d-1)V}
  \sum_p\frac{(\hat{p}^2)^2-\hat{p}^4}{(\hat{p}^2+M^2)^2}\,,
\end{align}
where
\begin{equation}
  \hat{p}^r\equiv \sum_\mu \hat{p}_\mu^r\,.
\end{equation}
Then
\begin{equation}
  \widetilde{\triangle}_{\mu\nu}(0)=\frac{w^2}{\frac12 \kappa^2 dw^2
    +\overline{H}_2-d\overline{H}_3}\left\{d\delta_{\mu\nu}-1\right\}
  +\frac{w^2}{\left[\frac12 \kappa^2 d w^2+\overline{H}_2
      -\overline{H}_1^2/\overline{H}_0\right]}\,,
\end{equation}
and so
\begin{equation}
  \widetilde{\triangle}(0)+\frac{H_\mu(0)H_\nu(0)}{H(0)^2}
  \widetilde{\triangle}_{\mu\nu}(0)
  =\frac{w^2}{\overline{H}_0}+\frac{w^2\overline{H}_1^2}{\overline{H}_0^2
    \left[\frac12 \kappa^2 dw^2+\overline{H}_2
      -\overline{H}_1^2/\overline{H}_0\right]}\,.
\end{equation}
Noting
\begin{align}
  \overline{H}_1 &=w-M^2\overline{H}_0\,,
  \\
  \overline{H}_2 &=1-2wM^2+M^4\overline{H}_0\,,
\end{align}
we get
\begin{equation}
  N\gR=  \frac{2}{M^2\overline{H}_0}
  +\frac{2\left(M^2\overline{H}_0-w\right)^2}
  {M^2\overline{H}_0^2\left[1+\frac12 d\kappa^2w^2-w^2/\overline{H}_0\right]}
  +\cO(1/N)\,.
  \label{gRx}
\end{equation}

\subsubsection{Lattice artifacts}
\label{AppB2b}

In the rest of this section we restrict attention to $d=2$
and consider large physical volumes.

To see the structure of artifacts we expand $w$ and $\overline{H}_0$:
\begin{equation} \label{wL}
  w=\frac{1}{4\pi}{\mathcal L}+\frac{M^2}{32\pi}(1-{\mathcal L})+\dots,
\end{equation}
\begin{equation}\label{mHL}
  M^2\overline{H}_0=\frac{1}{4\pi}-\frac{M^2}{32\pi}(2-{\mathcal L})\dots,
\end{equation}
where ${\mathcal L}=-\ln(M^2/32)$.

Taking the continuum limit 
we have $\kappa^2 w^2 = q^2 + \cO(1/L^2)$, with
$q$ given by eq.~\eqref{rqdef}.
Using eqs.~\eqref{wL}, \eqref{mHL} in eq.~\eqref{gRx} (with $d=2$)
we derive eq.~\eqref{gRart}.

\subsubsection{Numerical evaluation of \boldmath $\gR$}
\label{AppB2c}

In the infinite volume limit the gap equations are
\begin{equation} \label{Jmw}
  J(M) = w\,,
\end{equation}
and
\begin{equation} \label{kw}
  \kappa^2 \left( w - \frac{1}{f}\right) + M^2 = \frac{1}{w} \,,
\end{equation}
where
\begin{equation}
  J(M) = \int_{-\pi}^\pi \frac{\mathrm{d}^2p}{(2\pi)^2} 
  \frac{1}{K(p) + M^2 }
  =  \int_{-\pi}^\pi \frac{\mathrm{d}p}{2\pi} 
  \frac{1}{\sqrt{\omega(\omega+4)}} \,.
\end{equation}
Inserting $w=J(M)$ into eq.~\eqref{Jmw} one can find numerically
$M=M(f,\kappa)$.

One also has
\begin{equation}
  \overline{H}_0(M) = \int_{-\pi}^\pi \frac{\mathrm{d}^2p}{(2\pi)^2} 
  \frac{1}{\left(K(p) + M^2 \right)^2}
  = \int_{-\pi}^\pi \frac{\mathrm{d}p}{2\pi} 
  \frac{\omega+2}{\left[\omega(\omega+4)\right]^{3/2}} \,.
\end{equation}
Inserting these expressions into eq.~\eqref{gRx} it is easy to calculate 
$N \gR(aM)$.

\end{appendix}

\end{document}